\documentclass[12pt]{article}
\usepackage{graphicx}
\usepackage{color} 
\usepackage{amssymb} 
\usepackage{lineno}
\usepackage{aas_macros}   

\makeatletter
\renewcommand{\@seccntformat}[1]{}
\makeatother

\begin{document}

\title{Asteroseismic signatures of the helium-core flash}
\author{M. M. Miller Bertolami$^*$, T. Battich, A. H. C\'orsico,\\
  J. Christensen-Dalsgaard, L. G. Althaus }
\date{August 2019}
\maketitle

\section{Letter to {\em Nature Astronomy}}


{\bf All evolved stars with masses $M_\star\lesssim 2M_\odot$ undergo
  a helium(He)-core flash at the end of their first stage as a giant
  star. Although theoretically predicted more than 50 years
  ago\cite{1962ApJ...136..158S,1967ZA.....67..420T}, this core-flash
  phase has yet to be observationally probed. We show here that
  gravity modes (g modes) stochastically excited by He-flash driven
  convection are able to reach the stellar surface, and induce
  periodic photometric variabilities in hot-subdwarf stars with
  amplitudes of the order of a few mmag. As such they can now be
  detected by space-based photometry with the Transiting Exoplanet
  Survey Satellite (TESS) in relatively bright stars (e.g. magnitudes
  $I_C\lesssim 13$)\cite{2015ApJ...809...77S}. The range of predicted
  periods spans from a few thousand seconds to tens of thousand
  seconds, depending on the details of the excitation region. In
  addition, we find that stochastically excited pulsations
  reproduce the pulsations observed in a couple of He-rich hot
  subdwarf stars. These stars, and in particular the future TESS
  target Feige\,46, are the most promising candidates to probe the
  He-core flash for the first time.  }

\begin{figure}
\centering
\includegraphics[width=11.5cm]{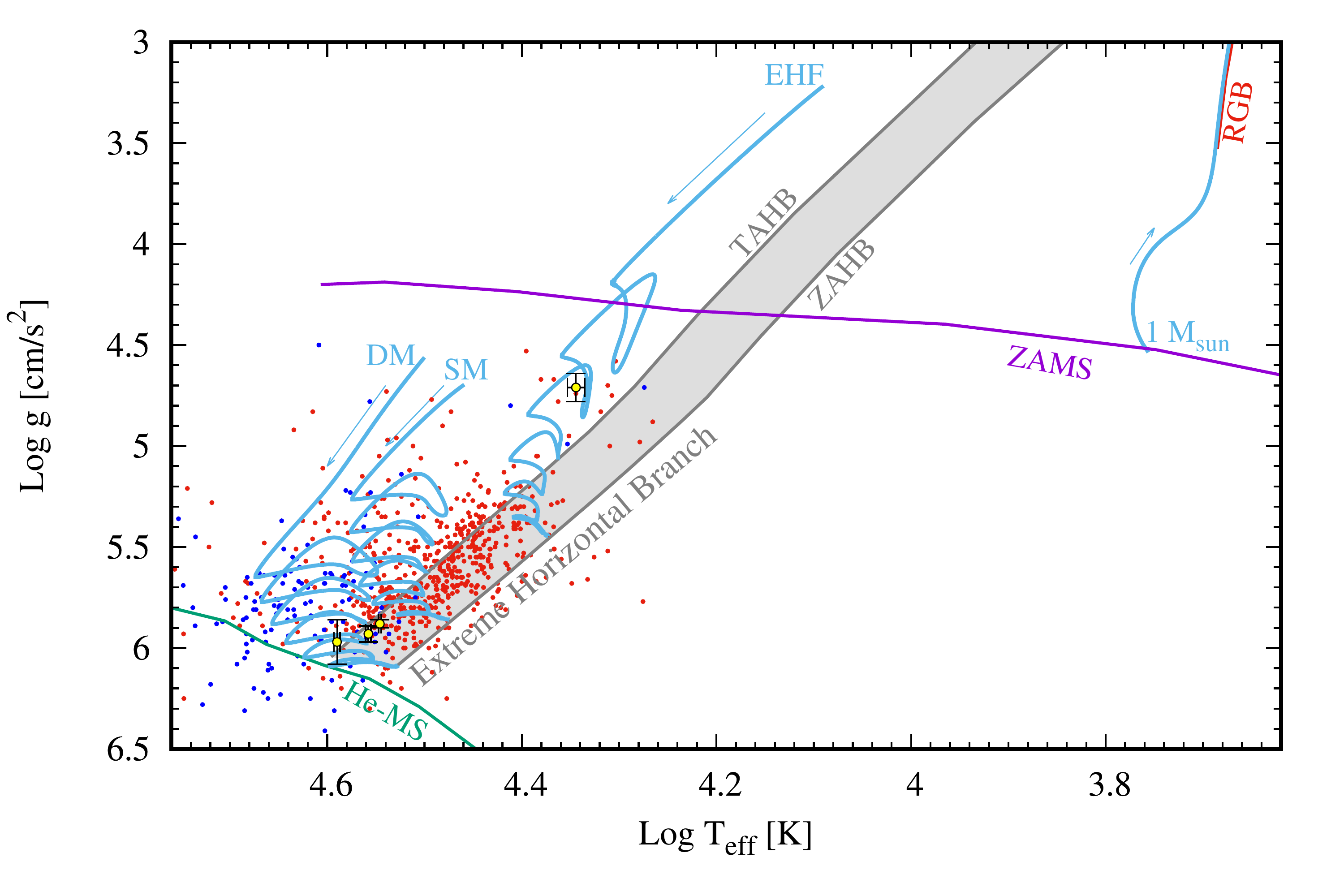}
\caption{{\bf Locus of hot subdwarfs in a $\log T_{\rm eff}-\log g$ diagram.} Red and blue dots indicate the positions of spectroscospically analysed hot H- and He-rich subdwarfs respectively\cite{2017A&A...602C...2G}. The He-rich hot-subdwarf pulsators studied in this work are shown as yellow circles with black errorbars. Errorbars correspond to the formal fitting errors reported by each author\cite{2015A&A...576A..65R,2012ApJ...753L..17O,2017MNRAS.465.3101J,feige46}. Hot subdwarf stars are located at the hottest end ($\log T_{\rm eff}\gtrsim 4.3$) of the horizontal branch (HB), bounded by the Zero Age Horizontal Branch (ZAHB) and Terminal Age Horizontal Branch (TAHB) shown with black lines\cite{1993ApJ...419..596D}.  For the sake of clarity the Zero Age Main Sequence of H-burning stars (ZAMS) and the Zero Age He Main Sequence formed by pure He-stars are shown with purple and green lines respectively\cite{2012sse..book.....K}. The red giant branch (RGB) is shown with a red line. The evolutionary tracks of model stars before the ZAHB (i.e. the pre-HB stage) are shown in blue\cite{2018A&A...614A.136B}. Blue labels indicate the location of our Early Hot-Flasher (EHF), Shallow Mixing (SM) and Deep Mixing (DM) models during the He subflashes (with initial abundances
  $(X_{\rm ZAMS},Y_{\rm ZAMS},Z_{\rm ZAMS})=(0.695,0.285,0.02)$).
    Also shown is the begining of the evolution of a $1 M_\odot$ main sequence star.  All theoretical sequences and tracks correspond to models with metallicity $Z_{\rm ZAMS}=0.02$. Arrows indicate the sense of evolution. }
 \label{fig1}
\end{figure}

\begin{figure}
\centering
\includegraphics[width=10.5cm]{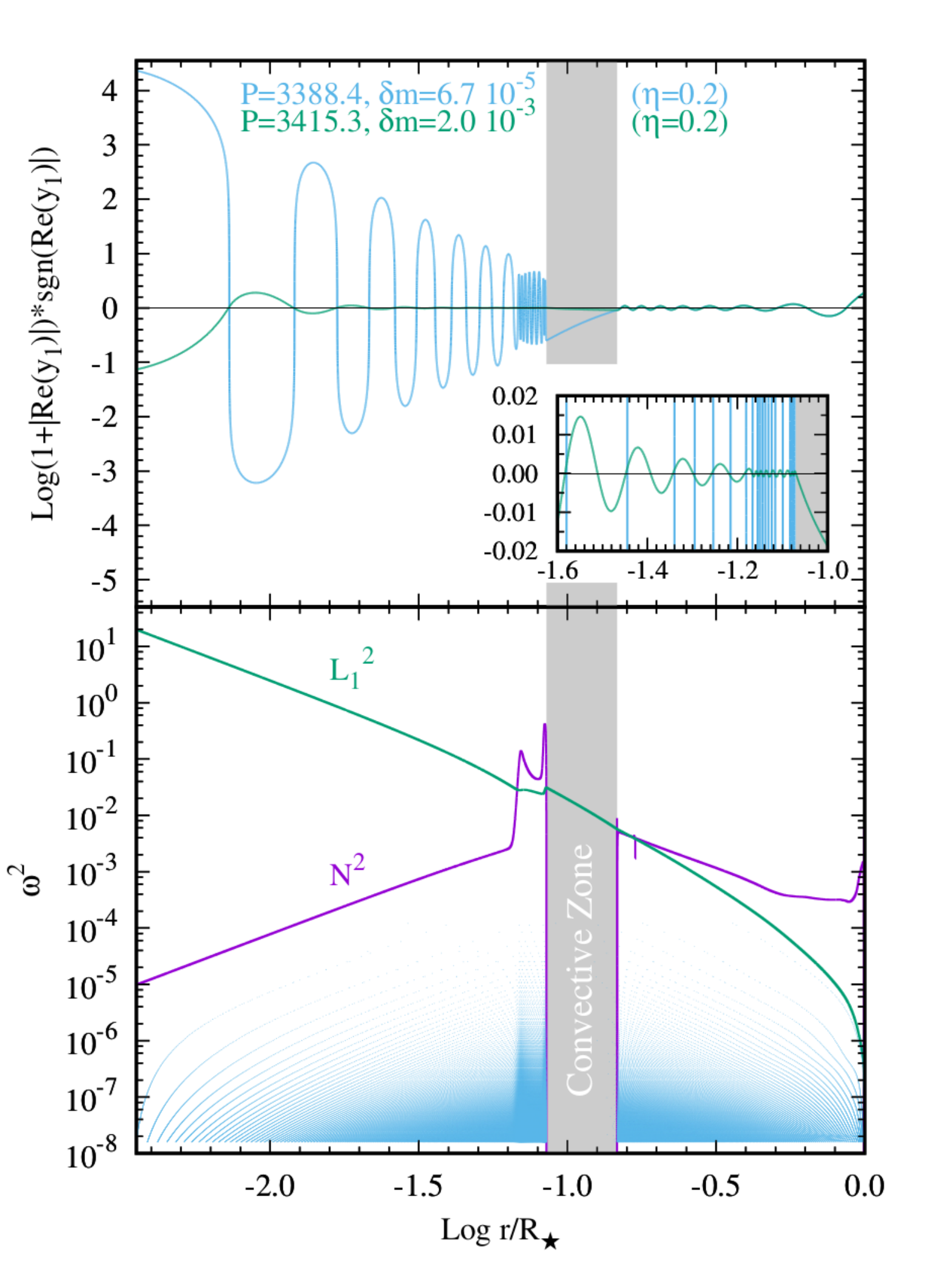}
\caption{{\bf Propagation diagram and pulsation eigenfunctions for $\ell=1$ modes in a  pre-EHB stellar model during a He subflash.} Lower panel: Squared Lamb (${L_1}^2$) and Brunt-Vais\"al\"a ($N^2$) frequencies together with the location of radial nodes (light-blue dots) for computed eigenfuctions at different angular eigenfrequencies $\omega^2$. Upper panel:  Magnitude of the radial displacement eigenfuctions\cite{1989nos..book.....U} $y_1$ of two consecutive radial orders and similar periods ($P$), but with very different global properties and predicted pulsation amplitudes ($\delta m$).  Note that eigenfunctions in the linear pulsation theory are arbitrarily set to $|y_1|=1$ at the surface and, consequently only relative differences are physically meaningful. The grey band displays the location of the convective zone driven by the He flash. The model shown in this figure corresponds to the maximum energy release by the He flash during the first subflash of the DM sequence shown in figure \ref{fig1}.}
 \label{fig2}
\end{figure}

After the end of their adult lives burning hydrogen (H) in the core,
low-mass stars (up to about 2\,$\rm M_\odot$) evolve into the Red
Giant Branch (RGB), where they develop a He core surrounded by a
H-burning shell. Due to the high densities, the structure of the
He-core is supported by the pressure of degenerate electrons. As H
burning proceeds, the mass of the He core increases and so does its
temperature. Eventually, this process leads to a violent ignition of
He in the core, which is termed the He-core flash.  Owing to the
neutrino cooling of the central regions, the initial ignition takes
place at some distance from the centre of the core.  This first He
flash, where the He-burning power reaches $L_{\rm He}\gtrsim 10^{10}
L_\odot$, is followed by a series of subflashes progressively deeper
in the core.  This phase lasts for about 2 Myr until degeneracy is
lifted and steady He-core burning can
start\cite{2012sse..book.....K}. He-core burning stars populate the
so-called Red Clump and the Horizontal Branch (HB) in the
Hertzsprung-Russell diagram\cite{2013osp..book.....C} (see figure
\ref{fig1}).

Due to the inherent difficulty of gathering direct information from
the core of stars, the He flash and its subflashes have remained
unprobed for almost 60 years. In the last decades, and with the advent
of space observatories like {\it Kepler}\cite{2011Sci...332..213C} and
{\it CoRoT}\cite{2010A&A...517A..22M}, the use of stellar pulsations to
learn about the interior of stars, a technique known as
asteroseismology, has grown to become one of the most flourishing
fields in stellar astrophysics. This is particularly true in the case
of solar-like (stochastically excited) pulsators. In these stars, the
bubbling motions in the convective envelope shake the star as a whole,
globally exciting a discrete spectrum of eigenmodes. The excited
eigenmodes are intrinsically damped by heat loss, so that the
pulsations reach an equilibrium amplitude in which the energy injected
by convective motions is balanced by the energy lost as heat by each
mode.  The unprecedented quality of the lightcurves from space
missions has allowed the identification of a large number of modes, in
solar-like pulsators.  While most pulsations observed in solar-like
pulsators are pressure modes (p modes), where the main restoring force
are pressure gradients, some solar-like pulsators exhibit stochastically excited
gravity modes (g modes)\cite{2011Sci...332..205B}. Gravity modes are
oscillation modes where the primary restoring force is gravity,
through buoyancy. Gravity modes are usually able to penetrate deeper
in the star than p modes. As a consequence, g modes are extremely
useful to constrain the internal structure of stars, and it has been
speculated that the He-core flash could produce such signals in the
pulsation spectrum of RGB stars\cite{2012ApJ...744L...6B,
  2018A&A...620A..43D}.

\begin{figure}
\centering
\includegraphics[width=10.5cm]{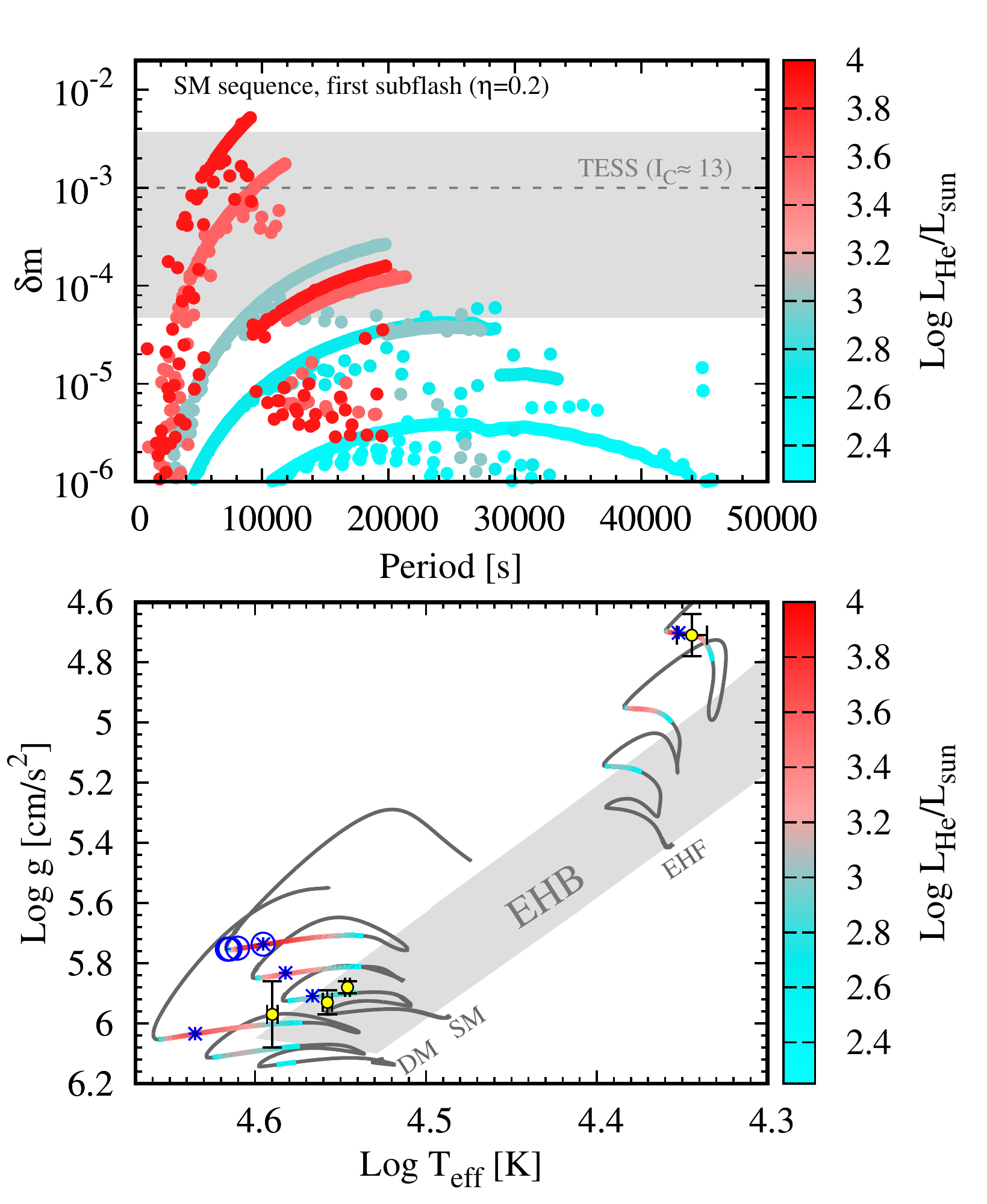}
\caption{ {\bf Evolution of $T_{\rm eff }$, $g$ and $L_{\rm He}$ in our stellar models and development of pulsations as compared with known He-rich subdwarf pulsators.} Lower panel: $T_{\rm eff }$-$g$ diagram of our computed DM, SM and EHF sequences for an initially He-enhanced population $(X_{\rm ZAMS},Y_{\rm ZAMS},Z_{\rm ZAMS})=(0.58,0.40,0.02))$  as compared with the known pulsating He-sdOBs shown as yellow points with black error bars\cite{2015A&A...576A..65R,2012ApJ...753L..17O,2017MNRAS.465.3101J,feige46} (from left to right UVO\,0825+15, Feige\,46, LS\,IV\,-14$^\circ$116, and KIC\,1718290). Errorbars correspond to the formal fitting errors reported by each author. Colours indicate the He-burning luminosity in the parts of the evolution in which models harbour an internal convective zone. Blue circles in the SM sequence indicate the models described in the upper panel and blue asterisks models described in figure 4 and in the supplementary material. Upper panel: Evolution of the predicted pulsation amplitudes of $\ell=1$ g modes during the development of the first subflash in the SM sequence shown in the lower panel (open blue circles). Colours indicate the He-burning luminosity in each model. The gray band indicates the typical  range of amplitudes observed in He-sdBs while the dash line indicates the expected sensitivity limit for TESS for a star with $I_{\rm C}\simeq 13$.}
 \label{fig3}
\end{figure}

We present here compelling evidence that g-mode pulsations can be
excited to observable levels by the inner convective zone driven by
He-core flashes. Previous work\cite{1990ApJ...363..694G} has shown
that the convective energy injected into p ($F_p$) and g modes ($F_g$)
scales as $ F_p \propto M_c^{15/2}\, F_c $ and $F_g \propto M_c\, F_c$, where $F_c$ is the total convective energy flux and
$M_c$ the Mach number of convective motions ---$M_c=v_c/c_{\rm
  sound}$, with $v_c$ the typical velocity of convective motions
and $c_{\rm sound}$ the sound speed. With values of $M_c\ll 1$, inner
convective zones are unable to excite p modes to observable
amplitudes, but the situation is more favourable for g modes. Still,
even in massive stars with intense core-H burning the total convective
flux in the core is only able to excite pulsations with very low
amplitudes ($10 \mu{\rm mag}$)\cite{2013MNRAS.430.1736S} because
oscillations are strongly damped in their massive radiative envelopes.
Fortunately, a much more
favourable situation takes place during the He flash of low-mass stars
that will populate the hot extreme of the horizontal branch (EHB,
$M^{\rm EHB}_\star\simeq 0.5 M_\odot$), known as hot
subdwarfs\cite{2016PASP..128h2001H} (sdO and sdB spectral types). In
these pre-EHB stars, the He-burning luminosity ($L_{\rm He}$) reaches
values of $L_{\rm He}\gtrsim 10^4 L_\odot$ during the subflashes, and
the mass of the radiative damping zone on top of the convective zone
is significantly smaller. Also, the mass of the star is one order of
magnitude lower and easier to disturb than that of more massive
stars. In addition, the sdBs and sdOs have calm, stable atmospheres
without convection or winds, making the surface imprint of non-radial
oscillations easier to detect.

We have performed numerical simulations to compute the predicted
amplitudes of stochastically driven $\ell=1$ g modes in pre-EHB stars
during the He subflashes in the framework of the hot-flasher scenario
\cite{1993ApJ...407..649C,2001ApJ...562..368B,2003ApJ...582L..43C,2004ApJ...602..342L,2008A&A...491..253M,2015Natur.523..318T}. Stellar
evolution models have been constructed with {\tt LPCODE} stellar
evolution code from initially $1M_\odot$ stars with different initial
He compositions\cite{2018A&A...614A.136B}. Within this scenario sdB
stars are formed when almost the whole H-rich envelope is stripped by
winds during the RGB phase, and the He-core flash develops once the
star has contracted away from the
RGB\cite{1993ApJ...407..649C}. Hot-flasher sequences are labeled as
``early hot flashers'' (EHF), ``shallow mixing'' (SM) or ``deep
mixing'' (DM) as in previous
works\cite{2004ApJ...602..342L,2008A&A...491..253M,
  2018A&A...614A.136B}. While in the case of the EHF flavour the sdB
model retains its original H/He atmosphere, in the SM and DM cases the
final surface composition is set by the action of mixing and burning
immediately after the main He-core flash, leading to a He-enriched
surface\cite{2008A&A...491..253M,2018A&A...614A.136B}.  Snapshots of
the internal structure of these stars during the subflashes in the
pre-EHB phase are then fed into the linear non-adiabatic pulsation
code {\tt LP-PUL}\cite{2006A&A...458..259C} for the computation of
normal modes (see figure \ref{fig2}), their kinetic energies ($E^{\rm
  linear}$) and the damping rates of the modes ($\gamma$) as
determined by all non-adiabatic effects. However, linear pulsation
analyses do not provide the actual amplitudes of the
pulsations. Amplitudes are obtained from balancing the power injected
by convection into g modes and the power lost due to non-adiabatic
effects during the oscillations (see Methods). The convective energy
flux $F_c$ was computed in the framework of the mixing-length theory
(MLT)\cite{2012sse..book.....K}, and the power spectrum of energy
supplied by convection to g modes computed under the simplifying
assumption that waves are excited by eddies and that the buoyancy
frequency ($N$) can be treated as discontinuous at convective
boundaries, as in previous work\cite{1990ApJ...363..694G,
  2013MNRAS.430.1736S} (see Methods). This model for gravity wave
excitation\cite{2013MNRAS.430.2363L} has been validated by numerical
studies\cite{2018JFM...854R...3C} and, despite important uncertainties
in the predicted energy spectrum, can be considered a conservative
estimation, as competing models predict larger gravity wave
excitation\cite{2013ApJ...772...21R}, and the consideration of a
smooth transition in the buoyancy frequency would also lead to an
enhancement in the predicted
excitation\cite{2013MNRAS.430.2363L}. Following previous
work\cite{2013MNRAS.430.1736S}, we parametrize the thickness of the
excitation region close to the convective boundaries by including a
parameter $\eta$, so that the geometrical size of the excitation
region is $\eta H_P$, where $H_P$ is the local pressure scale height
at the formal convective boundary. Values of the order of $\eta\sim
0.1$ are typical of convective boundary mixing under different
circumstances from the main sequence to the asymptotic giant
branch\cite{2016A&A...588A..25M}.

\begin{figure}
\centering
\includegraphics[width=10.5cm]{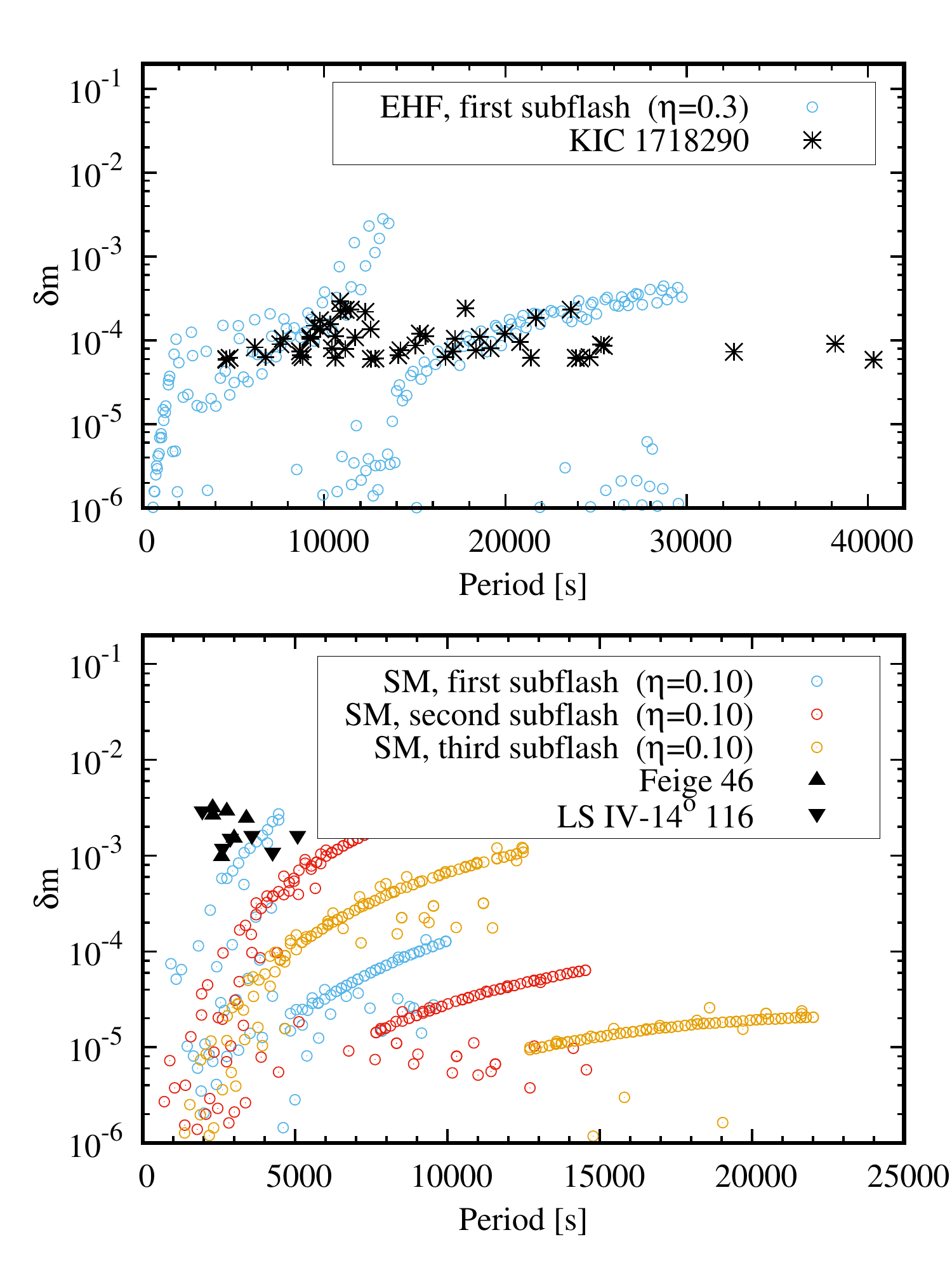}
\caption{{\bf Observed and predicted pulsation amplitudes of $\ell=1$ g modes.} 
Comparison of the predicted maximum pulsation amplitudes $|\delta m|$  with the actual periods and pulsation amplitudes observed in the He-rich subdwarfs. Predicted amplitudes are shown for the models indicated in figure \ref{fig3} with blue asterisks. Lower panel: Data from LS\,IV\,-14$^\circ$116, and  Feige\,46 and pulsation amplitudes predicted for our SM sequence ($Y_{\rm ZAMS}=0.4$) during the first three subflashes for $\eta=0.1$.  Upper panel: Data from  KIC\,1718290, and pulsation amplitudes predicted for our EHF sequence ($Y_{\rm ZAMS}=0.4$) during the first subflashes for $\eta=0.3$.}
 \label{fig4}
\end{figure}

Our simulations show that stochastically excited oscillations can
produce lightcurve variations of more than 1 mmag and even up to 10
mmag. As shown in figures \ref{fig3} and \ref{fig4} for the case of
our $Y_{\rm ZAMS}=0.4$ models, this is similar to the pulsations
detected in He-sdOBs stars and also well within the detectability
range of TESS for bright subdwarf stars\cite{2015ApJ...809...77S},
e.g. with magnitudes $I_C\lesssim 13$, of which more than one hundred
are known\cite{2017A&A...602C...2G}. The predicted amplitudes are even
larger in the case of our Y=0.285 sequences, for which the He-flashes
are more intense (see supplementary material).  In latter, less
intense flashes, the excitation becomes less intense and amplitudes
will be undetectable with current technology (figures \ref{fig3} and
\ref{fig4}). The longest excited periods are also the most excited,
and correspond to the turn-over frequency of convective motions
($\omega_c(\eta)$) at the excitation region of size $\eta H_P$ (see
Methods). This is a natural consequence of the fact that the energy
transport by convection occurs mostly via the largest possible eddies.
Consequently, the thinner the excitation region, the higher
characteristic frequency by a factor $1/\eta$. This shifts the bulk of
the g-mode power input to higher frequencies, increasing the
amplitudes of high frequency g modes for thinner excitation
regions. On the other hand, the characteristic frequency
$\omega_c(\eta)$ beyond which no pulsations are excited is also
decreased by the same factor. Our computations show that not all
eigenmodes are excited to the same amplitude.  Eigenmodes that
oscillate mostly in the outer parts of the star have lower inertia and
are more easily excited by convection, and therefore have larger
amplitudes (see figure \ref{fig2}).
The range of predicted periods spans from a couple thousand seconds to
tens of thousand seconds depending on the details of the excitation
region and the compactness of the stellar model. This provides for the
first time an extremely exciting prospect to probe the He-core flash
and to explore the nature of convective boundary mixing in He
flashes by asteroseismological means.

Given the large number of known hot
subdwarfs\cite{2017A&A...602C...2G} ($N>5000$) and given that the
pre-horizontal branch evolution lasts for about 2Myr ($\sim 2$\% of
the total He-core burning lifetime) we may expect that some of
known hot subdwarfs are in the pre-HB evolution. In particular,
He-rich hot subdwarfs (He-sdO/He-sdB spectral types) are natural candidates, 
 as the low surface abundance of H suggests that these stars
are in a fast evolutionary stage, where H has not had enough time to
diffuse and form a pure H atmosphere\cite{2008A&A...491..253M,
  2010MNRAS.409..582N}.   
Pre-EHB model sequences spend $0.5\,{\rm to}\,1$\% of their pre-EHB stage having
high He-burning luminosities $ L_{\rm He}\gtrsim 10^3 L_\odot$ and
thus a similar fraction of stars should display detectable stochastic
oscillations, see figure \ref{fig3}. Notably,
four He-rich hot-subdwarfs (out of $\sim 500$ He-rich subdwarfs
known\cite{2017A&A...602C...2G}) are known to pulsate, the pulsation
mechanism being still matter of
debate\cite{2018A&A...614A.136B,2005A&A...437L..51A,2012ApJ...753L..17O,2017MNRAS.465.3101J,feige46,2019MNRAS.482..758S}.
Noteworthy, the locus of these stars in the $\log g - \log T_{\rm
  eff}$ diagram overlaps with that of pre-EHB model sequences during
the He subflashes, see figures \ref{fig1} and \ref{fig3}.

 In figure \ref{fig4} we compare the range of periods excited in our
 models by convective motions with periods detected in the known
 He-rich subdwarf pulsators.  Remarkably, our models show that
 stochastic excitation during the He-core flash is able to excite
 periods in the range observed in He-rich subdwarfs for excitation
 regions with sizes typical of overshooting zones in stellar evolution
 (i.e. $\eta\simeq 0.1 ... 0.3$).  The ranges of amplitudes and
 periods measured in the twin stars LS\,IV\,-14$^\circ$116 and
 Feige\,46 are well reproduced by our SM models with enhanced initial
 He abundances ($Y=0.4$) in which the excitation takes place in a thin
 region of size $\sim 0.1 H_P$ (see figure \ref{fig4}). This agreement
 is reinforced by the fact that this sequence also nicely reproduces
 the spectroscopical determinations of $\log T_{\rm eff}$ and $\log g$
 in these two stars (figure \ref{fig3}).  On the other hand the large
 number of long periods observed in KIC\,1718290 are also nicely
 reproduced by our EHF sequences with a thin excitation region
 ($\eta\sim 0.3$) which also nicely reproduces the $\log T_{\rm eff}$
 and $\log g$ values determined for this star. This is true for both
 our EHF sequences with $Y_{\rm ZAMS}=0.4$ and $Y_{\rm ZAMS}=0.25$,
 see figures \ref{fig1}, \ref{fig3}, \ref{fig4} and the supplementary
 material. Although the longest modes are not reproduced, this can be
 easily accomodated by a slightly larger choice of $\eta$.  On the
 contrary, the very long and isolated pulsation modes observed in UVO
 0825+15\cite{2017MNRAS.465.3101J} cannot be reproduced by our
 sequences. While a choice of $\eta\sim 1$ would lead to the
 excitation of those modes, such excitation would be accompanied with
 the excitation of a range of shorter periods (as those observed in
 KIC\,1718290) that are not observed in this star. In addition, a
 choice of $\eta\sim 1$ is incompatible with the choices made in our
 model, most importantly with the assumption of a discontinuous
 buoyancy frequency at convective boundaries, and with the use of the
 frozen convection approximation (see Methods).  Consequently, we
 conclude that the stochastic excitation of pulsations provides a
 natural explanation for the lightcurves observed in KIC\,1718290,
 LS\,IV\,-14$^\circ$116, and Feige\,46 all of which lack another
 plausible explanation, but that they are unable to explain the
 lightcurve observed in UVO\,0825+15.

Our computations show that new space observatories, such as TESS,
allow us to search for direct asteroseismological signatures of the
He-core flash phase in hot subdwarf stars. This is particularly true
in the case of Feige\,46 which has already been scheduled as TESS
target\cite{feige46}. One of the distinctive features of a star
evolving thorugh this very fast evolutionary stage would be a large
rate of period drift\cite{2018A&A...614A.136B} $\dot{P}\sim
10^{-7}$--$10^{-4}$s/s. The determination of
such large rates of period drift in a pulsating hot subdwarf would be a
smoking gun for the He-core flash phase.




%

\newpage

\section{Methods}


\subsection{La Plata stellar evolution code}

All stellar evolution models presented in this work have been computed
with {\tt LPCODE}, a one-dimensional stellar evolution code that has
been widely used for the computation of full evolutionary sequences
from the zero age main sequence to the white dwarf stage.  The last
version of {\tt LPCODE} includes a state-of-the-art treatment of
atomic, molecular and conductive opacities as well as a detailed
treatment of stellar winds and convective boundary
mixing\cite{2016A&A...588A..25M, 2018NatAs...2..580G,
  2018NatAs...2..784G}.

Models of He-rich hot subdwarfs were constructed within the
hot-flashers scenario as in our previous
 work\cite{2018A&A...614A.136B}.  Models were calculated from the
evolution of initially $1 M_\odot$ models on the Zero Age Main
Sequence (ZAMS) to the end of the RGB  where the He-core flash
develops. At the tip of the RGB we switched on artificially enhanced
mass loss, removing different amounts of envelope mass to obtain the
different types of hot-flasher sequences. For the purpose of this
work, the particular value of the mass at the ZAMS and the treatment
of mass loss at RGB are not relevant, since it is the total envelope mass at He
ignition that determines the behaviour of the hot flashers.

\subsection{La Plata stellar pulsation code}

Normal oscillation modes and theoretical pulsation periods have been
computed with the non-adiabatic version of {\tt LP-PUL}\cite{2006A&A...458..259C}. {\tt LP-PUL} is a linear,
non-radial, non-adiabatic stellar pulsation code, which is
coupled to the {\tt LPCODE}. The {\tt LP-PUL} was widely used in
  studies of pulsation properties of late stages of low-mass stars. In
  particular it was recently used by our group to study the excitation
  of pulsations by means of the $\epsilon$ mechanism in hot pre-HB
  stars\cite{2018A&A...614A.136B}. {\tt LP-PUL} allows as to compute
  the damping rates of the oscillation modes ($\gamma$) in the full
  non-adiabatic regime as well as the kinetic energies
  of the modes $E^{\rm linear}(\omega,\ell)$ under the standard normalization\cite{1989nos..book.....U} ($\xi_{R_\star}/{R_\star}=1$, see below).  Non-adiabatic
  computations in this work rely on the frozen-in convection
  approximation, in which the perturbation  to the convective flux is
  neglected, an approximation that breaks down for modes with periods similar or
  longer than the convective  turn-over timescale.

\subsection{Computation of stochastically excited pulsations}
To compute the predicted maximum luminosity variations due to the excited
modes, we follow the approach of previous
 work\cite{2013MNRAS.430.1736S}.  Specifically, we compute the
equilibrium energy $E^{\rm eq}_{mode}(\omega, \ell)$ of the excited
modes by equating the energy injected by convection and the energy
dissipated by non-adiabatic effects,
\begin{equation}
E^{\rm eq}_{mode}(\omega, \ell)=\frac{1}{2}\frac{{\rm d}\dot{E_g}
}{{\rm d}\ln \omega {\rm d}\ell} N(\omega, \ell)^{-1} \gamma^{-1}
\label{eq:eq_energy}
\end{equation} 
where $N(\omega, \ell) \simeq (n+\ell/2)(2\ell+1)$ is the number of
modes in a logarithmic bin in $\omega$ (for a given $\ell$), $\gamma$
is the damping rate of the mode due to non-adiabatic effects, and
${\rm d}\dot{E_g}/{\rm d}\ln \omega {\rm d}\ell$ is the power spectrum
of energy supplied by convective motions to the g modes.

We compute $\gamma$ within the non-adiabatic linear oscillation theory with {\tt LP-PUL}\cite{2006A&A...458..259C} and estimate the power spectrum supplied by convective motions under the assumption that the excitation occurs at the edge of the convective boundaries, under the approximation of a discontinuity in the bouyancy frequency at the radiative-convective transition, as in previous  work\cite{1990ApJ...363..694G, 2013MNRAS.430.1736S},
\begin{eqnarray}
\frac{{\rm d}\dot{E_g} }{{\rm d}\ln \omega {\rm d}\ell}&=&B M_c L_{\rm conv} \nonumber\\
&\times&\left(\frac{\omega}{\omega_c}\right)^{-13/2}\ell^2\left(\frac{\eta H_P}{r}\right)^{3}
\left(1+\ell\frac{\eta H_P}{r}\right)
\exp\left[-\left(\frac{\ell}{\ell_{\rm max}}\right)^2\right]
\label{power}
\end{eqnarray}
which is valid over the range $\omega\ge \omega_c=v_{\rm MLT}/(\eta
H_P)$, and $\ell\gtrsim 1$. $L_{\rm conv}$ is the total power carried
by convection, $H_P$ is the pressure scale height, $v_{\rm MLT}$ is
the convective velocity, $M_c=v_{\rm MLT}/c_{\rm sound}$ is the
convective Mach number, $\omega_c$ is the convective turn-over
frequency, $\ell_{\rm max}= (r/\eta H_P)(\omega/\omega_c)^{3/2}$, and
both $H_P$ and $r$ are evaluated at the convective
boundaries. Convective velocities and luminosities are taken as a mean
value in the excitation region of size $0.2 H_P$ next to the formal
convective boundary.  It is worth noting that, as the flash-driven
convective zone has two convective boundaries (figure \ref{fig2}), the
total energy received by each mode is given by the sum of the energies
provided by the two convective boundaries (${E_g}^{\rm inf}$ and
${E_g}^{\rm sup}$), i.e.
\begin{equation}
\frac{{\rm d}\dot{E_g}}{{\rm d}\ln \omega {\rm d}\ell}=
\frac{{\rm d}\dot{E_g}^{\rm inf}}{{\rm d}\ln \omega {\rm d}\ell}+
\frac{{\rm d}\dot{E_g}^{\rm sup}}{{\rm d}\ln \omega {\rm d}\ell}
\end{equation}
where both  
${\rm d}\ln\dot{E_g}^{\rm inf}/{\rm d}\ln \omega {\rm d}\ell$
 and
${\rm d}\ln\dot{E_g}^{\rm sup}/{\rm d}\ln \omega {\rm d}\ell$ are computed from equation \ref{power} with the corresponding value of $H_P$.

Convective velocities and luminosities were computed with the help of
the  mixing-length theory \cite{1953ZA.....32..135V} as
\begin{equation}
v_{\rm MLT}=\left(\alpha_{\rm MLT} \frac{acG}{3}\frac{m}{r^2\kappa\rho}
\frac{T^4}{P}\nabla_{\rm ad}(\nabla_{\rm rad}-\nabla)\right)^{1/3}
\end{equation}
and
\begin{equation}
L_{\rm conv}= \frac{16 a c \pi}{3}\frac{GmT^4}{\kappa P}(\nabla_{\rm rad}-\nabla)
\end{equation}
 where all the quantities have the same meaning as in the classic textbook by Kippenhahn et al.\cite{2012sse..book.....K}.

In the linear stellar oscillation theory, the amplitudes of the
eigenfunctions are arbitrary and are usually normalized by choosing
$\xi_r({R_\star})=\delta r(r=R_\star)=R_\star$. As a consequence, the
mode energies computed within the linear theory $E^{\rm
  linear}(\omega,\ell)$,
\begin{equation}
E^{\rm linear}(\omega,\ell)=\omega^2/2
\int_0^{M_\star} \left[{\xi_r}^2+\Lambda(\omega,\ell)^2{\xi_h}^2\right] \rho r^2 dr
\label{eq:linear_energy}
\end{equation}
differ from the equilbrium energy $E^{\rm eq}_{mode}(\omega, \ell)$ by
a constant $A(\omega, \ell)^2$. By comparing the energies computed in
eqs. \ref{eq:eq_energy} and \ref{eq:linear_energy} we can derive
$A(\omega, \ell)$ and normalize the linear eigenfunctions to obtain
realistic predictions for the temperature, pressure, flux and radial
perturbations ($\delta T$, $\delta P$, and $\delta r$).

Once the realistic surface perturbations were obtained, following the work by Dziembowski\cite{1977AcA....27..203D} we computed the instantaneous disk-averaged magnitude variations produced by the stochastically excited g modes as
\begin{equation}
\delta m=1.086\left(b_\ell\frac{\delta F(t,r,\theta_0,\phi_0)}{F}+(2b_\ell-c_\ell)\frac{\xi_r(t,r,\theta_0,\phi_0)}{R}\right)_{r=R_\star}
\label{eq:deltam}
\end{equation}
where in this expression the perturbations of flux $\delta F(t,r,\theta_0,\phi_0)$ and radius $\xi_r(t,r,\theta_0,\phi_0)$ depend of the direction to the observer ($\theta_0,\phi_0$) and time $t$. From equation \ref{eq:deltam}, we can then derive the amplitude of the lightcurve of each mode as
\begin{equation}
|\delta m|=1.086 \sqrt{(b_l\Re(y_6)-c_l \Re(y_1))^2+(b_l \Im(y_6))^2}\times A(\omega, \ell),
\label{theequation}
\end{equation}
where the factor $A(\omega, \ell)$ renormalizes the $y_1$ and $y_6$ coming from the {\tt LP-PUL} (which defines $y_1(R_\star)=1$\cite{1989nos..book.....U}) to their physical value. In equation \ref{theequation}, $\Re$ and $\Im$ denote the real and imaginary parts of the eigenfunctions $y_1$ and $y_6$, which are defined as
\begin{equation}
y_1=\xi_r/r,\, \, \, \, \, y_6=\delta L_r/L_r
\end{equation}
where $\delta L_r=4\pi r^2 (\delta F_r + 2 F_r \xi_r/r)$ and, under
the assumption of no convection in that particular region (which is
our case in the outer regions of a hot star) we have $L_r= 4\pi r^2 F^R_r$,
 and we can write 
\begin{equation}
\frac{\delta F_r}{F_r}=\frac{\delta L_r}{L_r}-2\frac{\xi_r}{r}=y_6-2y_1.
\end{equation}
Equation \ref{theequation} allows us to compute the maximum predicted
amplitude for each normal mode (i.e. for the most favourable
orientation $\theta_0,\phi_0$).

\section{Code availability}
The {\tt LPCODE} and {\tt LP-PUL} codes used in this paper are
available under request from M.M.M.B. and A.H.C., respectively. Note
that {\tt LPCODE} and {\tt LP-PUL} are not suitably written for public
distribution.
\section{Data availability}
The data that support the plots within this paper and other findings
of this study are available from the corresponding author upon reasonable request.

\newpage
\appendix
\section{Supplementary Information}

\newpage
\begin{thebibliography}{10}
\expandafter\ifx\csname url\endcsname\relax
  \def\url#1{\texttt{#1}}\fi
\expandafter\ifx\csname urlprefix\endcsname\relax\def\urlprefix{URL }\fi
\providecommand{\bibinfo}[2]{#2}
\providecommand{\eprint}[2][]{\url{#2}}

\bibitem{1962ApJ...136..158S}
\bibinfo{author}{{Schwarzschild}, M.} \& \bibinfo{author}{{H{\"a}rm}, R.}
\newblock \bibinfo{title}{{Red Giants of Population II. II.}}
\newblock \emph{\bibinfo{journal}{\apj}} \textbf{\bibinfo{volume}{136}},
  \bibinfo{pages}{158} (\bibinfo{year}{1962}).

\bibitem{1967ZA.....67..420T}
\bibinfo{author}{{Thomas}, H.-C.}
\newblock \bibinfo{title}{{Sternentwicklung VIII. Der Helium-Flash bei einem
  Stern von 1. 3 Sonnenmassen}}.
\newblock \emph{\bibinfo{journal}{\zap}} \textbf{\bibinfo{volume}{67}},
  \bibinfo{pages}{420} (\bibinfo{year}{1967}).

\bibitem{2015ApJ...809...77S}
\bibinfo{author}{{Sullivan}, P.~W.} \emph{et~al.}
\newblock \bibinfo{title}{{The Transiting Exoplanet Survey Satellite:
  Simulations of Planet Detections and Astrophysical False Positives}}.
\newblock \emph{\bibinfo{journal}{\apj}} \textbf{\bibinfo{volume}{809}},
  \bibinfo{pages}{77} (\bibinfo{year}{2015}).
  \newblock

\bibitem{2017A&A...602C...2G}
\bibinfo{author}{{Geier}, S.} \emph{et~al.}
\newblock \bibinfo{title}{{The catalogue of radial velocity variable hot
  subluminous stars from the MUCHFUSS project (Corrigendum)}}.
\newblock \emph{\bibinfo{journal}{\aap}} \textbf{\bibinfo{volume}{602}},
  \bibinfo{pages}{C2} (\bibinfo{year}{2017}).

\bibitem{2015A&A...576A..65R}
\bibinfo{author}{{Randall}, S.~K.}, \bibinfo{author}{{Bagnulo}, S.},
  \bibinfo{author}{{Ziegerer}, E.}, \bibinfo{author}{{Geier}, S.} \&
  \bibinfo{author}{{Fontaine}, G.}
\newblock \bibinfo{title}{{The enigmatic He-sdB pulsator LS
  IV-14$^\circ$116: new insights from the VLT}}.
\newblock \emph{\bibinfo{journal}{\aap}} \textbf{\bibinfo{volume}{576}},
  \bibinfo{pages}{A65} (\bibinfo{year}{2015}).
  \newblock

\bibitem{2012ApJ...753L..17O}
\bibinfo{author}{{{\O}stensen}, R.~H.} \emph{et~al.}
\newblock \bibinfo{title}{{KIC 1718290: A Helium-rich V1093-Her-like Pulsator
  on the Blue Horizontal Branch}}.
\newblock \emph{\bibinfo{journal}{\apjl}} \textbf{\bibinfo{volume}{753}},
  \bibinfo{pages}{L17} (\bibinfo{year}{2012}).
  \newblock

\bibitem{2017MNRAS.465.3101J}
\bibinfo{author}{{Jeffery}, C.~S.} \emph{et~al.}
\newblock \bibinfo{title}{{Discovery of a variable lead-rich hot subdwarf: UVO
  0825+15}}.
\newblock \emph{\bibinfo{journal}{\mnras}} \textbf{\bibinfo{volume}{465}},
  \bibinfo{pages}{3101--3124} (\bibinfo{year}{2017}).
  \newblock

\bibitem{feige46}
\bibinfo{author}{{Latour}, M.}, \bibinfo{author}{{Green}, E.~M.} \&
  \bibinfo{author}{{Fontaine}, G.}
\newblock \bibinfo{title}{{Discovery of a second pulsating intermediate
  helium-enriched sdOB star}}.
\newblock \emph{\bibinfo{journal}{\aap}} \textbf{\bibinfo{volume}{623}},
  \bibinfo{pages}{L12} (\bibinfo{year}{2019}).
  \newblock

\bibitem{1993ApJ...419..596D}
\bibinfo{author}{{Dorman}, B.}, \bibinfo{author}{{Rood}, R.~T.} \&
  \bibinfo{author}{{O'Connell}, R.~W.}
\newblock \bibinfo{title}{{Ultraviolet Radiation from Evolved Stellar
  Populations. I. Models}}.
\newblock \emph{\bibinfo{journal}{\apj}} \textbf{\bibinfo{volume}{419}},
  \bibinfo{pages}{596} (\bibinfo{year}{1993}).
  \newblock

\bibitem{2012sse..book.....K}
\bibinfo{author}{{Kippenhahn}, R.}, \bibinfo{author}{{Weigert}, A.} \&
  \bibinfo{author}{{Weiss}, A.}
\newblock \emph{\bibinfo{title}{{Stellar Structure and Evolution}}}
  (\bibinfo{year}{2012}).

\bibitem{2018A&A...614A.136B}
\bibinfo{author}{{Battich}, T.}, \bibinfo{author}{{Miller Bertolami}, M.~M.},
  \bibinfo{author}{{C{\'o}rsico}, A.~H.} \& \bibinfo{author}{{Althaus}, L.~G.}
\newblock \bibinfo{title}{{Pulsational instabilities driven by the $\epsilon$
  mechanism in hot pre-horizontal branch stars. I. The hot-flasher scenario}}.
\newblock \emph{\bibinfo{journal}{\aap}} \textbf{\bibinfo{volume}{614}},
  \bibinfo{pages}{A136} (\bibinfo{year}{2018}).
  \newblock

\bibitem{1989nos..book.....U}
\bibinfo{author}{{Unno}, W.}, \bibinfo{author}{{Osaki}, Y.},
  \bibinfo{author}{{Ando}, H.}, \bibinfo{author}{{Saio}, H.} \&
  \bibinfo{author}{{Shibahashi}, H.}
\newblock \emph{\bibinfo{title}{{Nonradial oscillations of stars}}}
  (\bibinfo{year}{1989}).

\bibitem{2013osp..book.....C}
\bibinfo{author}{{Cassisi}, S.} \& \bibinfo{author}{{Salaris}, M.}
\newblock \emph{\bibinfo{title}{{Old Stellar Populations: How to Study the
  Fossil Record of Galaxy Formation}}} (\bibinfo{year}{2013}).

\bibitem{2011Sci...332..213C}
\bibinfo{author}{{Chaplin}, W.~J.} \emph{et~al.}
\newblock \bibinfo{title}{{Ensemble Asteroseismology of Solar-Type Stars with
  the NASA Kepler Mission}}.
\newblock \emph{\bibinfo{journal}{Science}} \textbf{\bibinfo{volume}{332}},
  \bibinfo{pages}{213} (\bibinfo{year}{2011}).
  \newblock

\bibitem{2010A&A...517A..22M}
\bibinfo{author}{{Mosser}, B.} \emph{et~al.}
\newblock \bibinfo{title}{{Red-giant seismic properties analyzed with CoRoT}}.
\newblock \emph{\bibinfo{journal}{\aap}} \textbf{\bibinfo{volume}{517}},
  \bibinfo{pages}{A22} (\bibinfo{year}{2010}).
  \newblock

\bibitem{2011Sci...332..205B}
\bibinfo{author}{{Beck}, P.~G.} \emph{et~al.}
\newblock \bibinfo{title}{{Kepler Detected Gravity-Mode Period Spacings in a
  Red Giant Star}}.
\newblock \emph{\bibinfo{journal}{Science}} \textbf{\bibinfo{volume}{332}},
  \bibinfo{pages}{205} (\bibinfo{year}{2011}).

\bibitem{2012ApJ...744L...6B}
\bibinfo{author}{{Bildsten}, L.}, \bibinfo{author}{{Paxton}, B.},
  \bibinfo{author}{{Moore}, K.} \& \bibinfo{author}{{Macias}, P.~J.}
\newblock \bibinfo{title}{{Acoustic Signatures of the Helium Core Flash}}.
\newblock \emph{\bibinfo{journal}{\apjl}} \textbf{\bibinfo{volume}{744}},
  \bibinfo{pages}{L6} (\bibinfo{year}{2012}).
  \newblock

\bibitem{2018A&A...620A..43D}
\bibinfo{author}{{Deheuvels}, S.} \& \bibinfo{author}{{Belkacem}, K.}
\newblock \bibinfo{title}{{Seismic characterization of red giants going through
  the helium-core flash}}.
\newblock \emph{\bibinfo{journal}{\aap}} \textbf{\bibinfo{volume}{620}},
  \bibinfo{pages}{A43} (\bibinfo{year}{2018}).
  \newblock

\bibitem{1990ApJ...363..694G}
\bibinfo{author}{{Goldreich}, P.} \& \bibinfo{author}{{Kumar}, P.}
\newblock \bibinfo{title}{{Wave generation by turbulent convection}}.
\newblock \emph{\bibinfo{journal}{\apj}} \textbf{\bibinfo{volume}{363}},
  \bibinfo{pages}{694--704} (\bibinfo{year}{1990}).

\bibitem{2013MNRAS.430.1736S}
\bibinfo{author}{{Shiode}, J.~H.}, \bibinfo{author}{{Quataert}, E.},
  \bibinfo{author}{{Cantiello}, M.} \& \bibinfo{author}{{Bildsten}, L.}
\newblock \bibinfo{title}{{The observational signatures of convectively excited
  gravity modes in main-sequence stars}}.
\newblock \emph{\bibinfo{journal}{\mnras}} \textbf{\bibinfo{volume}{430}},
  \bibinfo{pages}{1736--1745} (\bibinfo{year}{2013}).
  \newblock

\bibitem{2016PASP..128h2001H}
\bibinfo{author}{{Heber}, U.}
\newblock \bibinfo{title}{{Hot Subluminous Stars}}.
\newblock \emph{\bibinfo{journal}{\pasp}} \textbf{\bibinfo{volume}{128}},
  \bibinfo{pages}{082001} (\bibinfo{year}{2016}).
  \newblock

\bibitem{1993ApJ...407..649C}
\bibinfo{author}{{Castellani}, M.} \& \bibinfo{author}{{Castellani}, V.}
\newblock \bibinfo{title}{{Mass Loss in Globular Cluster Red Giants: an
  Evolutionary Investigation}}.
\newblock \emph{\bibinfo{journal}{\apj}} \textbf{\bibinfo{volume}{407}},
  \bibinfo{pages}{649} (\bibinfo{year}{1993}).

\bibitem{2001ApJ...562..368B}
\bibinfo{author}{{Brown}, T.~M.}, \bibinfo{author}{{Sweigart}, A.~V.},
  \bibinfo{author}{{Lanz}, T.}, \bibinfo{author}{{Landsman}, W.~B.} \&
  \bibinfo{author}{{Hubeny}, I.}
\newblock \bibinfo{title}{{Flash Mixing on the White Dwarf Cooling Curve:
  Understanding Hot Horizontal Branch Anomalies in NGC 2808}}.
\newblock \emph{\bibinfo{journal}{\apj}} \textbf{\bibinfo{volume}{562}},
  \bibinfo{pages}{368--393} (\bibinfo{year}{2001}).
  \newblock

\bibitem{2003ApJ...582L..43C}
\bibinfo{author}{{Cassisi}, S.}, \bibinfo{author}{{Schlattl}, H.},
  \bibinfo{author}{{Salaris}, M.} \& \bibinfo{author}{{Weiss}, A.}
\newblock \bibinfo{title}{{First Full Evolutionary Computation of the Helium
  Flash-induced Mixing in Population II Stars}}.
\newblock \emph{\bibinfo{journal}{\apjl}} \textbf{\bibinfo{volume}{582}},
  \bibinfo{pages}{L43--L46} (\bibinfo{year}{2003}).
  \newblock

\bibitem{2004ApJ...602..342L}
\bibinfo{author}{{Lanz}, T.}, \bibinfo{author}{{Brown}, T.~M.},
  \bibinfo{author}{{Sweigart}, A.~V.}, \bibinfo{author}{{Hubeny}, I.} \&
  \bibinfo{author}{{Landsman}, W.~B.}
\newblock \bibinfo{title}{{Flash Mixing on the White Dwarf Cooling Curve: Far
  Ultraviolet Spectroscopic Explorer Observations of Three He-rich sdB Stars}}.
\newblock \emph{\bibinfo{journal}{\apj}} \textbf{\bibinfo{volume}{602}},
  \bibinfo{pages}{342--355} (\bibinfo{year}{2004}).
  \newblock

\bibitem{2008A&A...491..253M}
\bibinfo{author}{{Miller Bertolami}, M.~M.}, \bibinfo{author}{{Althaus},
  L.~G.}, \bibinfo{author}{{Unglaub}, K.} \& \bibinfo{author}{{Weiss}, A.}
\newblock \bibinfo{title}{{Modeling He-rich subdwarfs through the hot-flasher
  scenario}}.
\newblock \emph{\bibinfo{journal}{\aap}} \textbf{\bibinfo{volume}{491}},
  \bibinfo{pages}{253--265} (\bibinfo{year}{2008}).
  \newblock

\bibitem{2015Natur.523..318T}
\bibinfo{author}{{Tailo}, M.} \emph{et~al.}
\newblock \bibinfo{title}{{Rapidly rotating second-generation progenitors for
  the `blue hook' stars of {$\omega$} Centauri}}.
\newblock \emph{\bibinfo{journal}{\nat}} \textbf{\bibinfo{volume}{523}},
  \bibinfo{pages}{318--321} (\bibinfo{year}{2015}).
  \newblock

\bibitem{2006A&A...458..259C}
\bibinfo{author}{{C{\'o}rsico}, A.~H.}, \bibinfo{author}{{Althaus}, L.~G.} \&
  \bibinfo{author}{{Miller Bertolami}, M.~M.}
\newblock \bibinfo{title}{{New nonadiabatic pulsation computations on full PG
  1159 evolutionary models: the theoretical GW Virginis instability strip
  revisited}}.
\newblock \emph{\bibinfo{journal}{\aap}} \textbf{\bibinfo{volume}{458}},
  \bibinfo{pages}{259--267} (\bibinfo{year}{2006}).
  \newblock

\bibitem{2013MNRAS.430.2363L}
\bibinfo{author}{{Lecoanet}, D.} \& \bibinfo{author}{{Quataert}, E.}
\newblock \bibinfo{title}{{Internal gravity wave excitation by turbulent
  convection}}.
\newblock \emph{\bibinfo{journal}{\mnras}} \textbf{\bibinfo{volume}{430}},
  \bibinfo{pages}{2363--2376} (\bibinfo{year}{2013}).
  \newblock

\bibitem{2018JFM...854R...3C}
\bibinfo{author}{{Couston}, L.-A.}, \bibinfo{author}{{Lecoanet}, D.},
  \bibinfo{author}{{Favier}, B.} \& \bibinfo{author}{{Le Bars}, M.}
\newblock \bibinfo{title}{{The energy flux spectrum of internal waves generated
  by turbulent convection}}.
\newblock \emph{\bibinfo{journal}{Journal of Fluid Mechanics}}
  \textbf{\bibinfo{volume}{854}}, \bibinfo{pages}{R3} (\bibinfo{year}{2018}).
  \newblock

\bibitem{2013ApJ...772...21R}
\bibinfo{author}{{Rogers}, T.~M.}, \bibinfo{author}{{Lin}, D.~N.~C.},
  \bibinfo{author}{{McElwaine}, J.~N.} \& \bibinfo{author}{{Lau}, H.~H.~B.}
\newblock \bibinfo{title}{{Internal Gravity Waves in Massive Stars: Angular
  Momentum Transport}}.
\newblock \emph{\bibinfo{journal}{\apj}} \textbf{\bibinfo{volume}{772}},
  \bibinfo{pages}{21} (\bibinfo{year}{2013}).
  \newblock

\bibitem{2016A&A...588A..25M}
\bibinfo{author}{{Miller Bertolami}, M.~M.}
\newblock \bibinfo{title}{{New models for the evolution of post-asymptotic
  giant branch stars and central stars of planetary nebulae}}.
\newblock \emph{\bibinfo{journal}{\aap}} \textbf{\bibinfo{volume}{588}},
  \bibinfo{pages}{A25} (\bibinfo{year}{2016}).
  \newblock

\bibitem{2010MNRAS.409..582N}
\bibinfo{author}{{Naslim}, N.}, \bibinfo{author}{{Jeffery}, C.~S.},
  \bibinfo{author}{{Ahmad}, A.}, \bibinfo{author}{{Behara}, N.~T.} \&
  \bibinfo{author}{{{\c S}ah{\`i}n}, T.}
\newblock \bibinfo{title}{{Abundance analyses of helium-rich subluminous B
  stars}}.
\newblock \emph{\bibinfo{journal}{\mnras}} \textbf{\bibinfo{volume}{409}},
  \bibinfo{pages}{582--590} (\bibinfo{year}{2010}).
  \newblock

\bibitem{2005A&A...437L..51A}
\bibinfo{author}{{Ahmad}, A.} \& \bibinfo{author}{{Jeffery}, C.~S.}
\newblock \bibinfo{title}{{Discovery of pulsation in a helium-rich subdwarf B
  star}}.
\newblock \emph{\bibinfo{journal}{\aap}} \textbf{\bibinfo{volume}{437}},
  \bibinfo{pages}{L51--L54} (\bibinfo{year}{2005}).

\bibitem{2019MNRAS.482..758S}
\bibinfo{author}{{Saio}, H.} \& \bibinfo{author}{{Jeffery}, C.~S.}
\newblock \bibinfo{title}{{The excitation of g-mode pulsations in hot
  helium-rich subdwarfs}}.
\newblock \emph{\bibinfo{journal}{\mnras}} \textbf{\bibinfo{volume}{482}},
  \bibinfo{pages}{758--761} (\bibinfo{year}{2019}).

\bibitem{2018NatAs...2..580G}
\bibinfo{author}{{Gesicki}, K.}, \bibinfo{author}{{Zijlstra}, A.~A.} \&
  \bibinfo{author}{{Miller Bertolami}, M.~M.}
\newblock \bibinfo{title}{{The mysterious age invariance of the planetary
  nebula luminosity function bright cut-off}}.
\newblock \emph{\bibinfo{journal}{Nature Astronomy}}
  \textbf{\bibinfo{volume}{2}}, \bibinfo{pages}{580--584}
  (\bibinfo{year}{2018}).
  \newblock

\bibitem{2018NatAs...2..784G}
\bibinfo{author}{{Guerrero}, M.~A.} \emph{et~al.}
\newblock \bibinfo{title}{{The inside-out planetary nebula around a born-again
  star}}.
\newblock \emph{\bibinfo{journal}{Nature Astronomy}}
  \textbf{\bibinfo{volume}{2}}, \bibinfo{pages}{784--789}
  (\bibinfo{year}{2018}).
  \newblock

\bibitem{1953ZA.....32..135V}
\bibinfo{author}{{Vitense}, E.}
\newblock \bibinfo{title}{{Die Wasserstoffkonvektionszone der Sonne. Mit 11
  Textabbildungen}}.
\newblock \emph{\bibinfo{journal}{\zap}} \textbf{\bibinfo{volume}{32}},
  \bibinfo{pages}{135} (\bibinfo{year}{1953}).

\bibitem{1977AcA....27..203D}
\bibinfo{author}{{Dziembowski}, W.}
\newblock \bibinfo{title}{{Light and radial velocity variations in a
  nonradially oscillating star}}.
\newblock \emph{\bibinfo{journal}{\actaa}} \textbf{\bibinfo{volume}{27}},
  \bibinfo{pages}{203--211} (\bibinfo{year}{1977}).




\bibitem{2018A&A...614A.136B}
\bibinfo{author}{{Battich}, T.}, \bibinfo{author}{{Miller Bertolami}, M.~M.},
  \bibinfo{author}{{C{\'o}rsico}, A.~H.} \& \bibinfo{author}{{Althaus}, L.~G.}
\newblock \bibinfo{title}{{Pulsational instabilities driven by the $\epsilon$
  mechanism in hot pre-horizontal branch stars. I. The hot-flasher scenario}}.
\newblock \emph{\bibinfo{journal}{\aap}} \textbf{\bibinfo{volume}{614}},
  \bibinfo{pages}{A136} (\bibinfo{year}{2018}).
\newblock.

\bibitem{2015A&A...576A..65R}
\bibinfo{author}{{Randall}, S.~K.}, \bibinfo{author}{{Bagnulo}, S.},
  \bibinfo{author}{{Ziegerer}, E.}, \bibinfo{author}{{Geier}, S.} \&
  \bibinfo{author}{{Fontaine}, G.}
\newblock \bibinfo{title}{{The enigmatic He-sdB pulsator LS
  IV-14$^\circ$116: new insights from the VLT}}.
\newblock \emph{\bibinfo{journal}{\aap}} \textbf{\bibinfo{volume}{576}},
  \bibinfo{pages}{A65} (\bibinfo{year}{2015}).
  \newblock

\bibitem{2012ApJ...753L..17O}
\bibinfo{author}{{{\O}stensen}, R.~H.} \emph{et~al.}
\newblock \bibinfo{title}{{KIC 1718290: A Helium-rich V1093-Her-like Pulsator
  on the Blue Horizontal Branch}}.
\newblock \emph{\bibinfo{journal}{\apjl}} \textbf{\bibinfo{volume}{753}},
  \bibinfo{pages}{L17} (\bibinfo{year}{2012}).
  \newblock

\bibitem{2017MNRAS.465.3101J}
\bibinfo{author}{{Jeffery}, C.~S.} \emph{et~al.}
\newblock \bibinfo{title}{{Discovery of a variable lead-rich hot subdwarf: UVO
  0825+15}}.
\newblock \emph{\bibinfo{journal}{\mnras}} \textbf{\bibinfo{volume}{465}},
  \bibinfo{pages}{3101--3124} (\bibinfo{year}{2017}).
  \newblock

\bibitem{feige46}
\bibinfo{author}{{Latour}, M.}, \bibinfo{author}{{Green}, E.~M.} \&
  \bibinfo{author}{{Fontaine}, G.}
\newblock \bibinfo{title}{{Discovery of a second pulsating intermediate
  helium-enriched sdOB star}}.
\newblock \emph{\bibinfo{journal}{\aap}} \textbf{\bibinfo{volume}{623}},
  \bibinfo{pages}{L12} (\bibinfo{year}{2019}).
  \newblock
  
\bibitem{1989nos..book.....U}
\bibinfo{author}{{Unno}, W.}, \bibinfo{author}{{Osaki}, Y.},
  \bibinfo{author}{{Ando}, H.}, \bibinfo{author}{{Saio}, H.} \&
  \bibinfo{author}{{Shibahashi}, H.}
\newblock \emph{\bibinfo{title}{{Nonradial oscillations of stars}}}
  (\bibinfo{year}{1989}).


  
\end{thebibliography}

\subsection{Sequences with $Y_{\rm ZAMS}=0.285$}
\begin{figure}
\centering
\includegraphics[width=9cm]{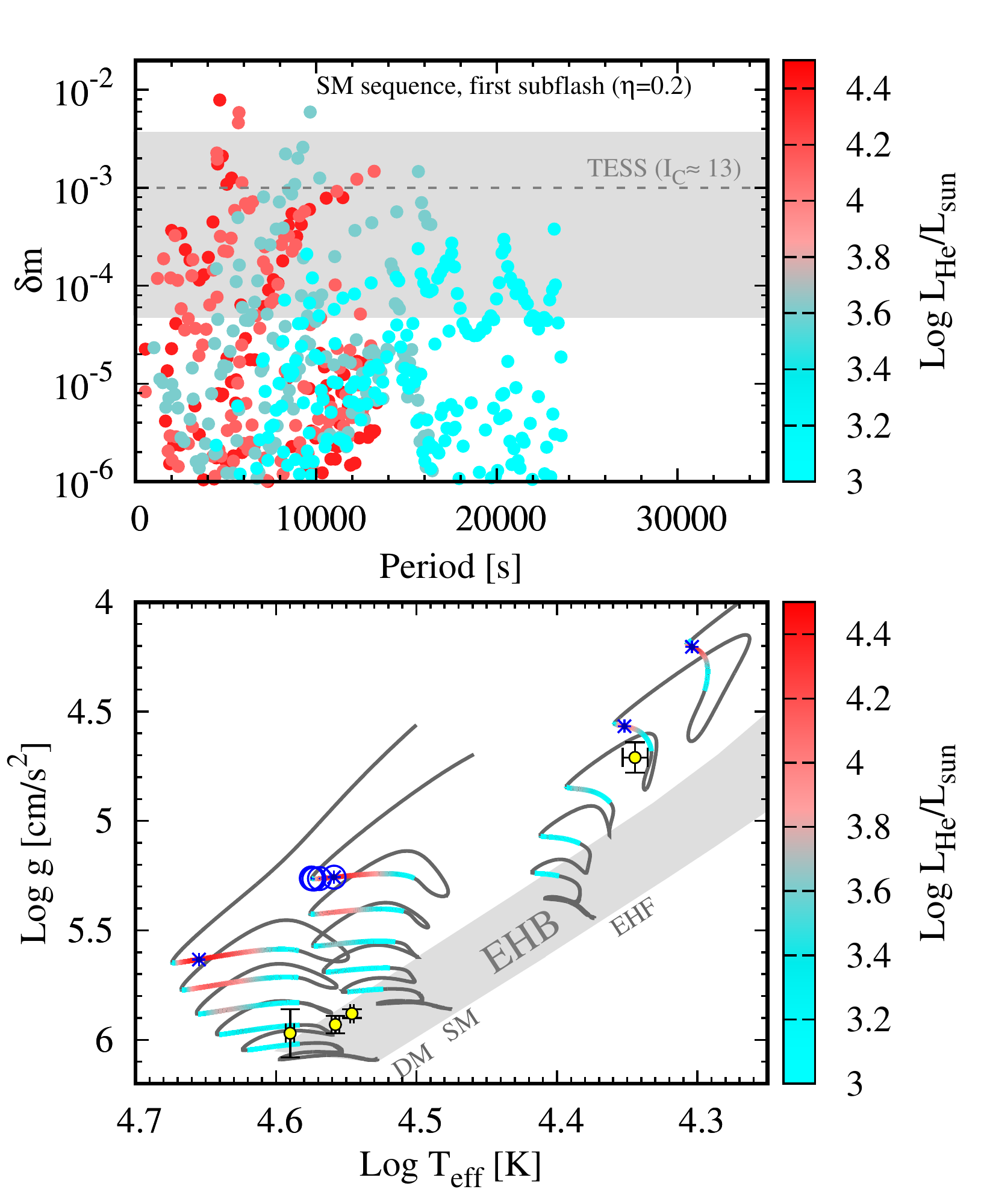}
\caption{{\bf Evolution of $T_{\rm eff }$, $g$ and $L_{\rm He}$ in our stellar models and development of pulsations as compared with known He-rich subdwarf pulsators.} Lower panel: $T_{\rm eff }$-$g$ diagram of our computed DM, SM and EHF sequences for an initially He-enhanced population $(X_{\rm ZAMS},Y_{\rm ZAMS},Z_{\rm ZAMS})=(0.695,0.285,0.02))$  as compared with the known pulsating He-sdOBs shown as yellow points with black error bars (from left to right UVO\,0825+15, Feige\,46, LS\,IV\,-14$^\circ$116, and KIC\,1718290). Errorbars correspond to the formal fitting errors provided by each author\cite{2015A&A...576A..65R,2012ApJ...753L..17O,2017MNRAS.465.3101J,feige46}. Colours indicate the He-burning luminosity in the parts of the evolution in which models harbour an internal convective zone. Blue circles in the SM sequence indicate the models described in the upper panel and blue asterisks models described in figure 4 of the main text and in figures \ref{Supp_fig4bis}, \ref{Supp_fig4UVO}, and \ref{Supp_fig1}. Upper panel: Evolution of the predicted pulsation amplitudes of $\ell=1$ g modes during the development of the first subflash in the SM sequence shown in the lower panel (open blue circles). Colours indicate the He-burning luminosity in each model. The gray band indicates the typical  range of amplitudes observed in He-sdBs while the dash line indicates the expected sensitivity limit for TESS for a star with $I_{\rm C}\simeq 13$.}
 \label{Supp_fig3bis}
\end{figure}

\begin{figure}
\centering
\includegraphics[width=9cm]{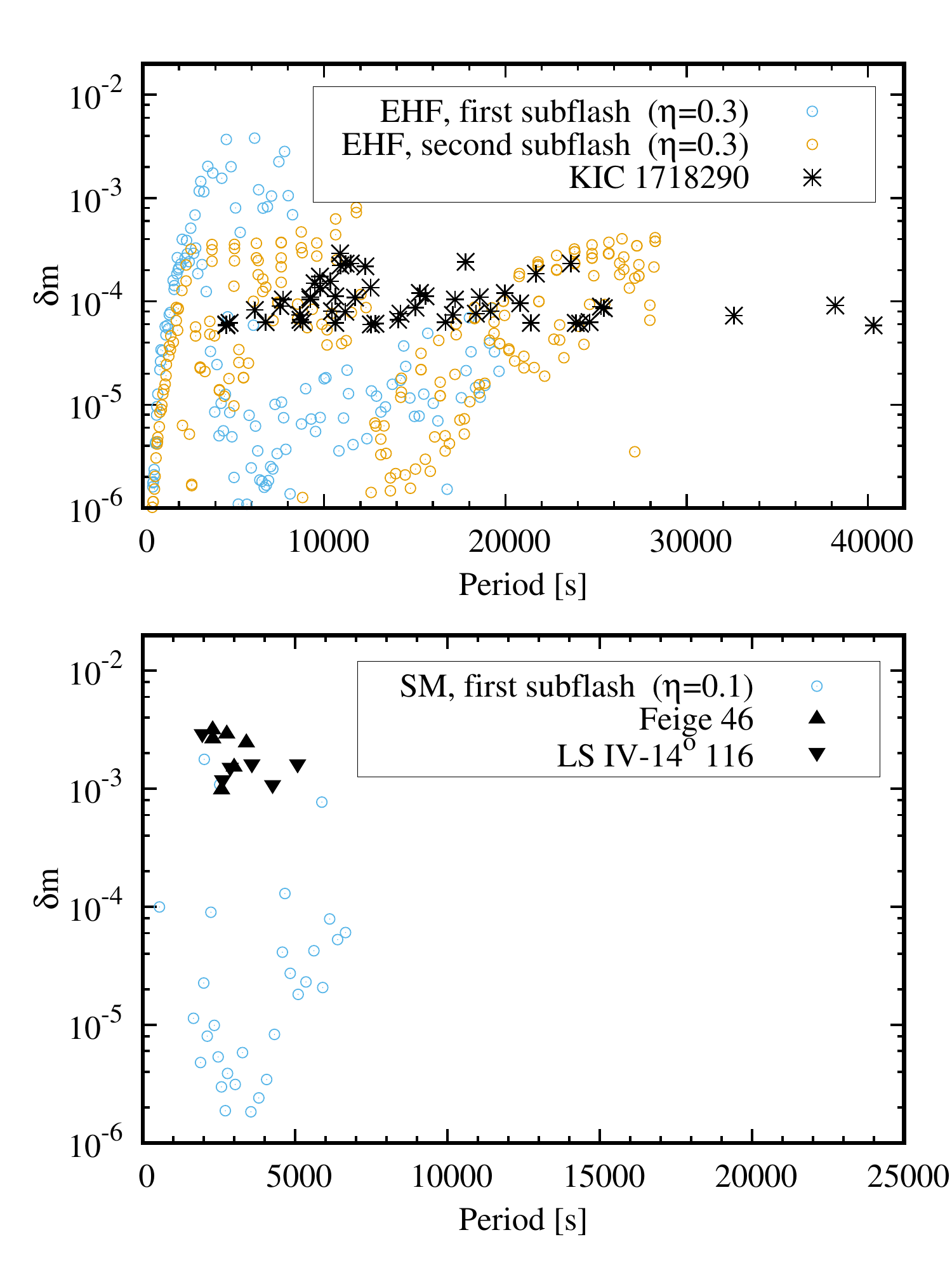}
\caption{{\bf Observed and predicted pulsation amplitudes of $\ell=1$ g modes.} 
Comparison of the predicted maximum pulsation amplitudes $|\delta m|$  with the actual periods and pulsation amplitudes observed in the He-rich subdwarfs. Predicted amplitudes are shown for the models indicated in figure 3 of the main text with blue asterisks. Lower panel: Data from LS\,IV\,-14$^\circ$116, and  Feige\,46 and pulsation amplitudes predicted for our SM sequence ($Y_{\rm ZAMS}=0.285$) during the first three subflashes for $\eta=0.1$.  Upper panel: Data from  KIC\,1718290, and pulsation amplitudes predicted for our EHF sequence ($Y_{\rm ZAMS}=0.285$) during the first subflashes for $\eta=0.3$.}
 \label{Supp_fig4bis}
\end{figure}

Supplementary figures \ref{Supp_fig3bis} and \ref{Supp_fig4bis} show
the same information as figures 3 and 4 of the main text but for our
sequences with a standard initial He abundance of $Y_{\rm
  ZAMS}=0.285$. This sequences harbour larger He cores, and as a
consequence undergo more intense He subflashes than the $Y_{\rm
  ZAMS}=0.4$ sequences of figures 3 and 4 in the main text. Note the
difference in the range of the colour bars in figure 3 of the main
text and supplementary figure \ref{Supp_fig3bis} here. As a
consequence these sequences also predict larger amplitudes due to
stochastic excitation at the convective boundaries. However, as
already noted by Battich et al.\cite{2018A&A...614A.136B} these
sequences do evolve at lower surface gravities due to their higher
luminosity and, consequently the SM and DM tracks do not overlap
during the first thermal pulses with the location of
LS\,IV\,-14$^\circ$116, and UVO\,0825+15 (supplementary figure
\ref{Supp_fig3bis}), and are consequently not favoured as models for
these stars. On the contrary the $Y_{\rm ZAMS}=0.285$ EHF sequence
does go thorugh the location of KIC\,1718290 and, in particular
overlaps its location during the second He subflash. Interestingly
during this second flash the models predict the excitation of g modes
in the right range of periods and amplitudes (supplementary figure
\ref{Supp_fig4bis}). We conclude that this sequence offers also a
viable model for KIC\,1718290 and its observed behaviour.

\newpage
\subsection{UVO\,0825+15}
\begin{figure}[h!]
\centering
\includegraphics[width=9cm]{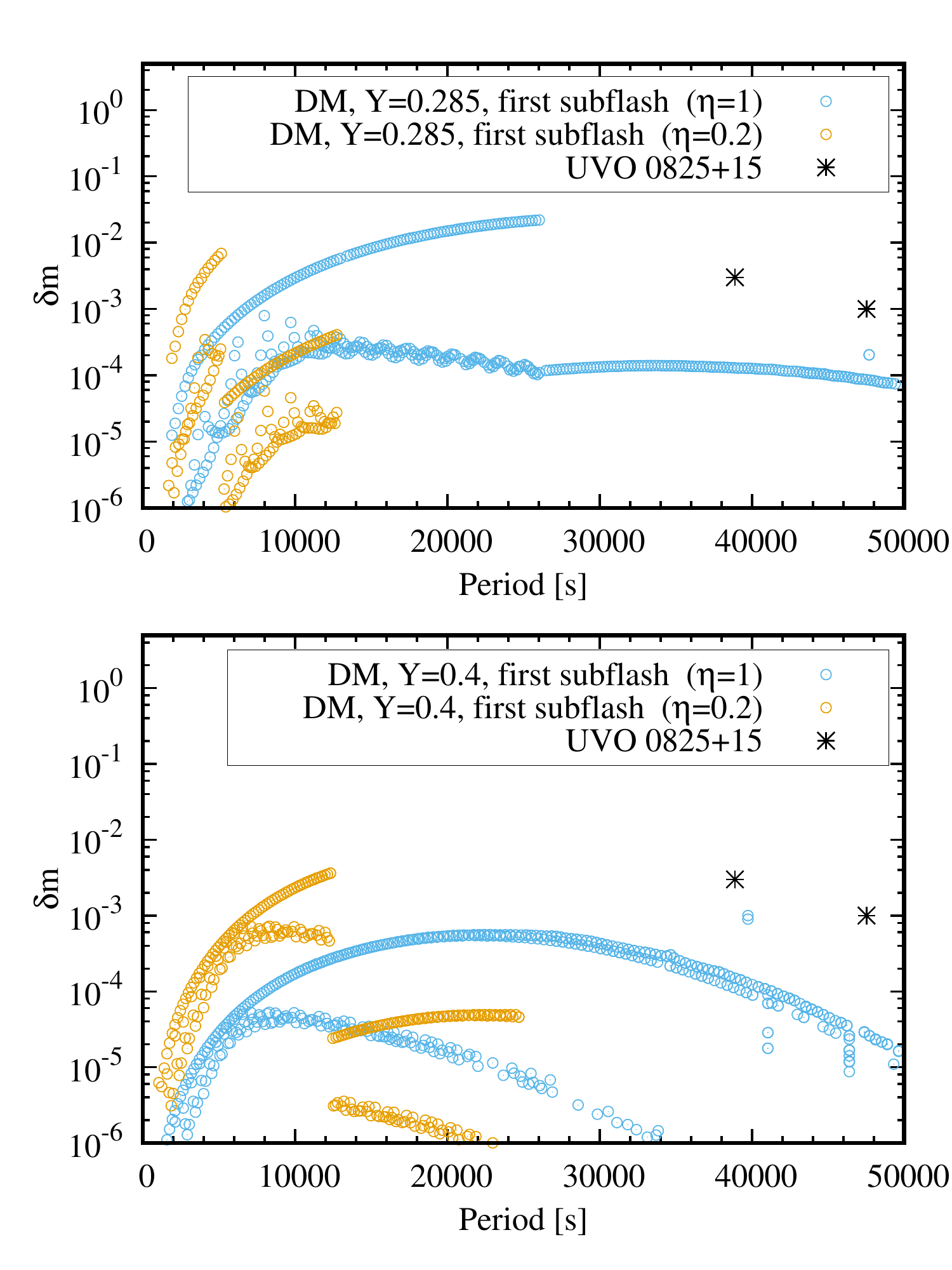}
\caption{{\bf Observed  pulsation amplitudes observed in  UVO\,0825+15 as compared with the predicted amplitudes of $\ell=1$ g modes in our DM sequences.} 
 Lower panel: Comparison with our $Y_{\rm ZAMS}=0.4$ DM sequence for two different values of $\eta$. Upper panel: Comparison with our $Y_{\rm ZAMS}=0.285$ DM sequence for two different values of $\eta$.}
 \label{Supp_fig4UVO}
\end{figure}

Supplementary figure \ref{Supp_fig4UVO} shows the observed periods in  UVO\,0825+15 compared with the excited modes in our DM sequences with $Y_{\rm ZAMS}=0.285$ (supplementary figure \ref{Supp_fig3bis}) and with $Y_{\rm ZAMS}=0.4$ (figure 3 of the main text). As is apparent from the figure, our models fail to reproduce the observed periods. Even when they manage to excite periods up to $P\gtrsim 40000$s under the extreme assuption of $\eta=1$, such models also predict a large range of shorter periods which are not observed in the star. 

\newpage
\subsection{Mode inertia and excitation at each convective boundary.}

\begin{figure}
\centering
\includegraphics[width=9cm]{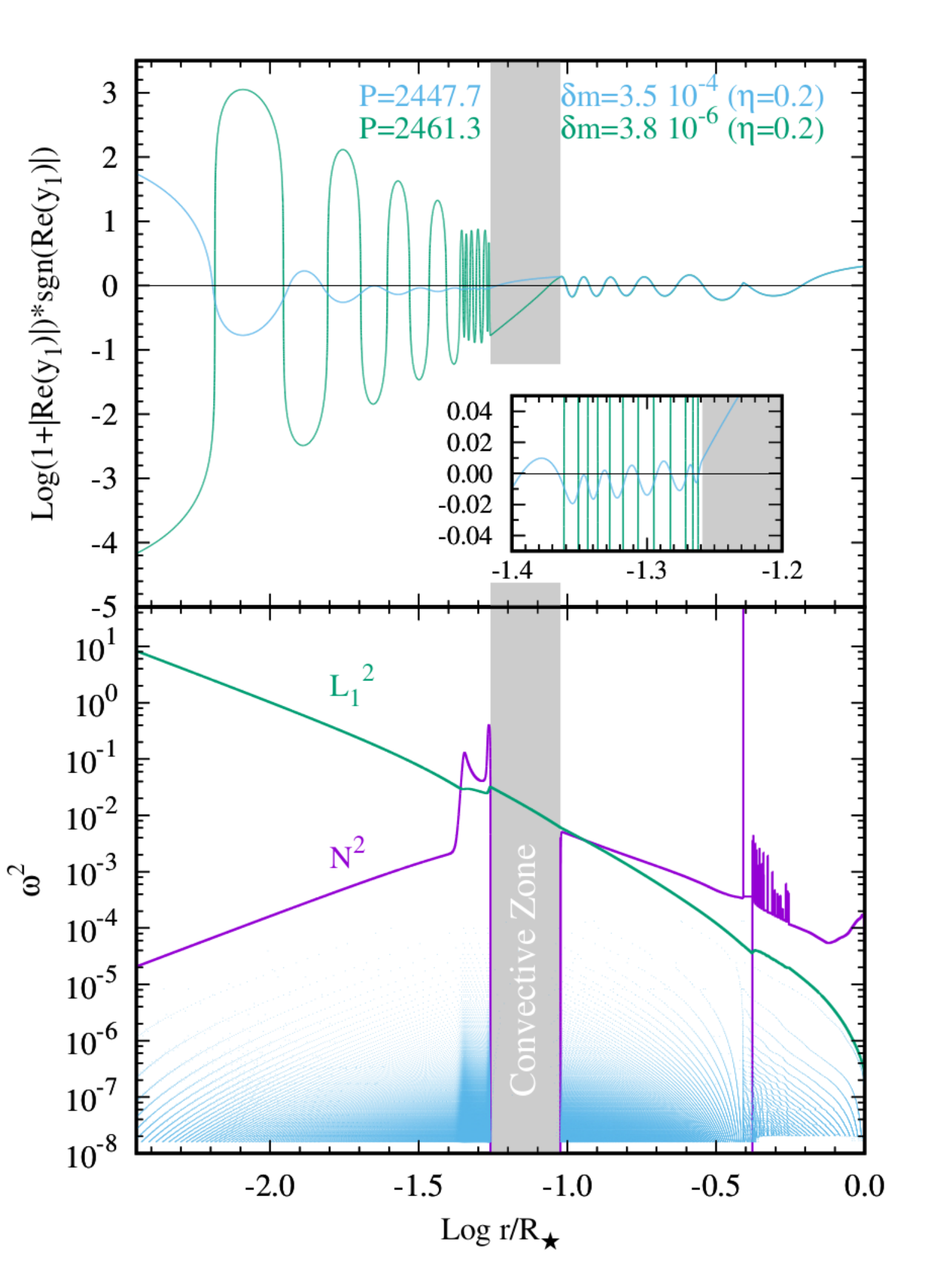}
\caption{{\bf Propagation diagram and pulsation eigenfunctions for $\ell=1$ modes in a  pre-EHB stellar model during a He subflash.} Lower panel: Squared Lamb (${L_1}^2$) and Brunt-Vais\"al\"a ($N^2$) frequencies together with the location of radial nodes (light-blue dots) for computed eigenfuctions at different angular eigenfrequencies $\omega^2$. Upper panel:  Magnitude of the radial displacement eigenfuctions\cite{1989nos..book.....U} $y_1$ of two consecutive radial ordersand similar periods ($P$), but with very different global properties and predicted pulsation amplitudes ($\delta m$).  Note that eigenfunctions in the linear pulsation theory are arbitrarily set to $|y_1|=1$ at the surface and, consequently only relative differences are physically meaningful. The grey band displays the location of the convective zone driven by the He flash. The model shown in this figure corresponds to the maximum energy release by the He flash during the first subflash of the SM sequence shown in figure 1 of the main text ($Y_{\rm ZAMS}=0.285$).}
 \label{Supp_fig2bis}
\end{figure}

The first row in supplementary figures \ref{Supp_fig1} and \ref{Supp_fig2} show the
maximum predicted amplitudes ($\delta m$) for stochastically excited
pulsations for DM, SM and EHF models (first, second and third columns
respectively) during the first subflash. As is apparent from these
plots, two main families of modes arise, one with amplitudes large
enough to be observed ($\delta m\gtrsim 10^{-4}$) and a second one
with very low amplitudes. The reason for this different behaviour can
be traced to the different eigenfunctions of the modes. While modes with large amplitudes have eigenfunctions that show similar amplitudes in the core and in the envelope of the star (see upper panels in figure 2 of the main text and supplementary figure \ref{Supp_fig2bis}), modes with smaller amplitudes at the surface have eigenfunctions with very large amplitudes in the core. This leads to very different mode inertias ($E^{\rm linear}$, second row in
supplementary figures \ref{Supp_fig1}, \ref{Supp_fig2}, \ref{Supp_fig11} and \ref{Supp_fig22})\footnote{Note that $E^{\rm
    linear}$ is a proxy for the inertia of the eigenmodes, as it
  indicates the total kinetic energy of the modes when the surface
  radial perturbation is $\xi_r({R_\star})=\delta
  r(r=R_\star)=R_\star$.} and damping rates (third row in supplementary figures
 \ref{Supp_fig1}, \ref{Supp_fig2}, \ref{Supp_fig11} and \ref{Supp_fig22}) of the respective
eigenmodes. Modes that are excited to larger amplitudes have
unnormalized energies, $E^{\rm linear}$, about five orders of magnitude
lower than low amplitude modes. These large inertia modes have,
however smaller damping rates $\gamma$ by about 1 to 2 orders of
magnitude (lower branch of modes in the third row of supplementary figures
\ref{Supp_fig1}, \ref{Supp_fig2}, \ref{Supp_fig11} and \ref{Supp_fig22}), which leads to equilibrium
energies (second row in supplementary figures \ref{Supp_fig1}, \ref{Supp_fig2}, \ref{Supp_fig11} and \ref{Supp_fig22})
only 1 to 2 orders of magnitude larger. In view of the large
difference in the inertia of the modes of about 5 orders of magnitude
these modes are excited to amplitudes much smaller than the low
inertia modes. 

Another interesting feature of our computations is that the lower
convective boundary of the convective shell is the one providing most
of the energy in our simulations (see second row of supplementary figures \ref{Supp_fig1}, \ref{Supp_fig2}, \ref{Supp_fig11} and \ref{Supp_fig22}). This is due to the smaller spatial
scales, which concentrates the energy injected into gravity waves and
higher frequencies and the higher turbulent velocities at this
convective boundary, larger Mach numbers, which lead to a larger total
energy flux transferred to gravity waves.

\begin{figure}
\includegraphics[width=14cm]{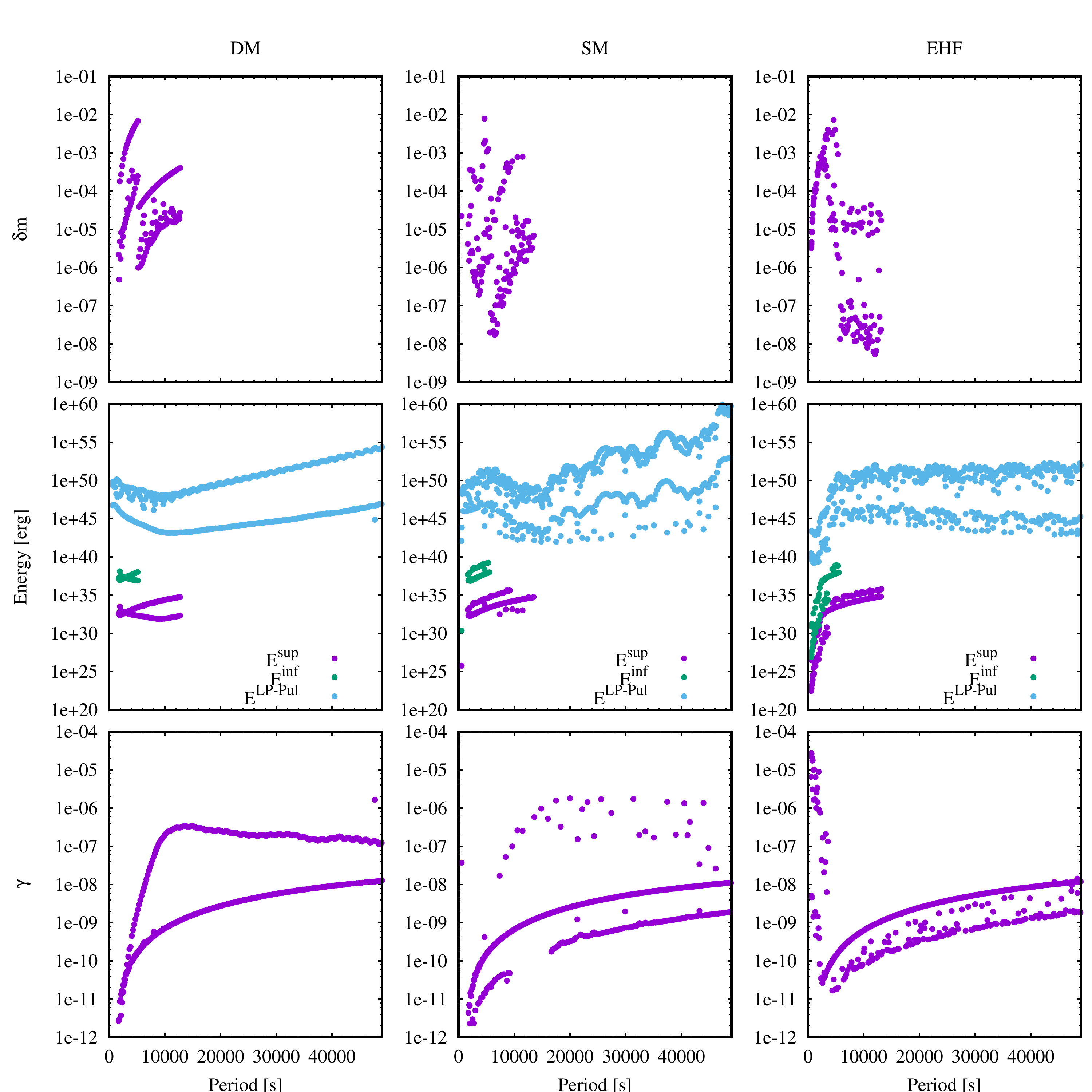}
\caption{{\bf Pulsations properties of DM, SM and EHF models with canonical initial abundances ($Y=0.285$) during the first subflash and under the assumption of $\eta=0.2$.} First, second, and third columns correspond to DM, SM and EHF models respectively. Top row shows the predicted pulsation amplitudes for eigenmodes at different periods. Middle row indicates the unnormalized energies, $E^{\rm linear}$, computed by the linear pulsation code ({\tt LP-PUL}) for all eigenmodes compared with the actual equilibrium energies derived from equation 1 (Methods section, main text) for the lower ($E^{\rm inf}$) and upper ($E^{\rm sup}$) convective boundaries of the He-flash driven convective zone. Bottom row shows the  damping rates ($\gamma$) computed with {\tt LP-PUL} for each eigenmode.}
 \label{Supp_fig1}
\end{figure}

\begin{figure}
\includegraphics[width=14cm]{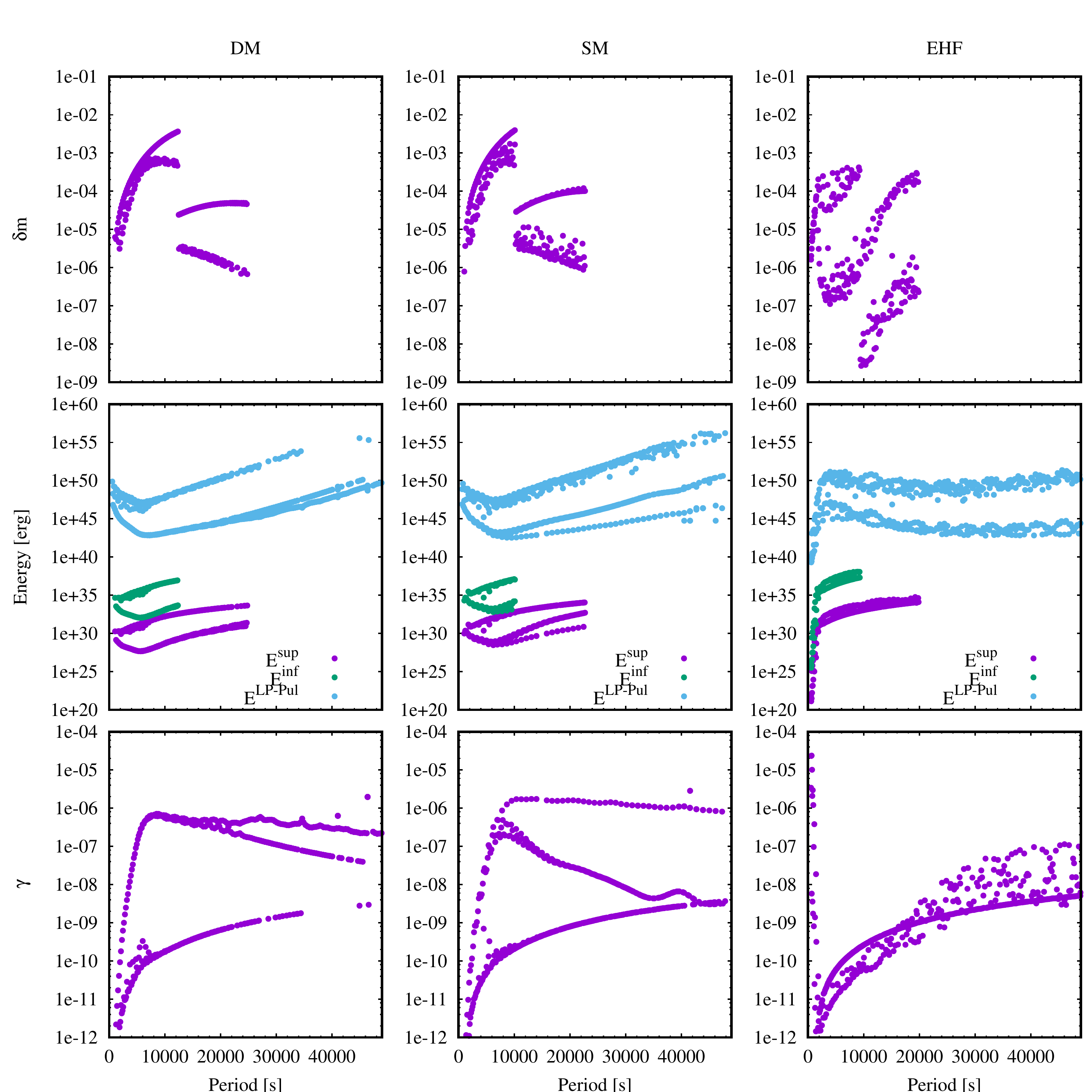}
\caption{{\bf Pulsations properties of DM, SM and EHF models with He-enriched  initial abundances ($Y=0.4$)  during the first subflash and under the assumption of $\eta=0.2$.} Rows and columns have the same meaning as in supplementary figure \ref{Supp_fig1}.}
 \label{Supp_fig2}
\end{figure}

\begin{figure}
\includegraphics[width=14cm]{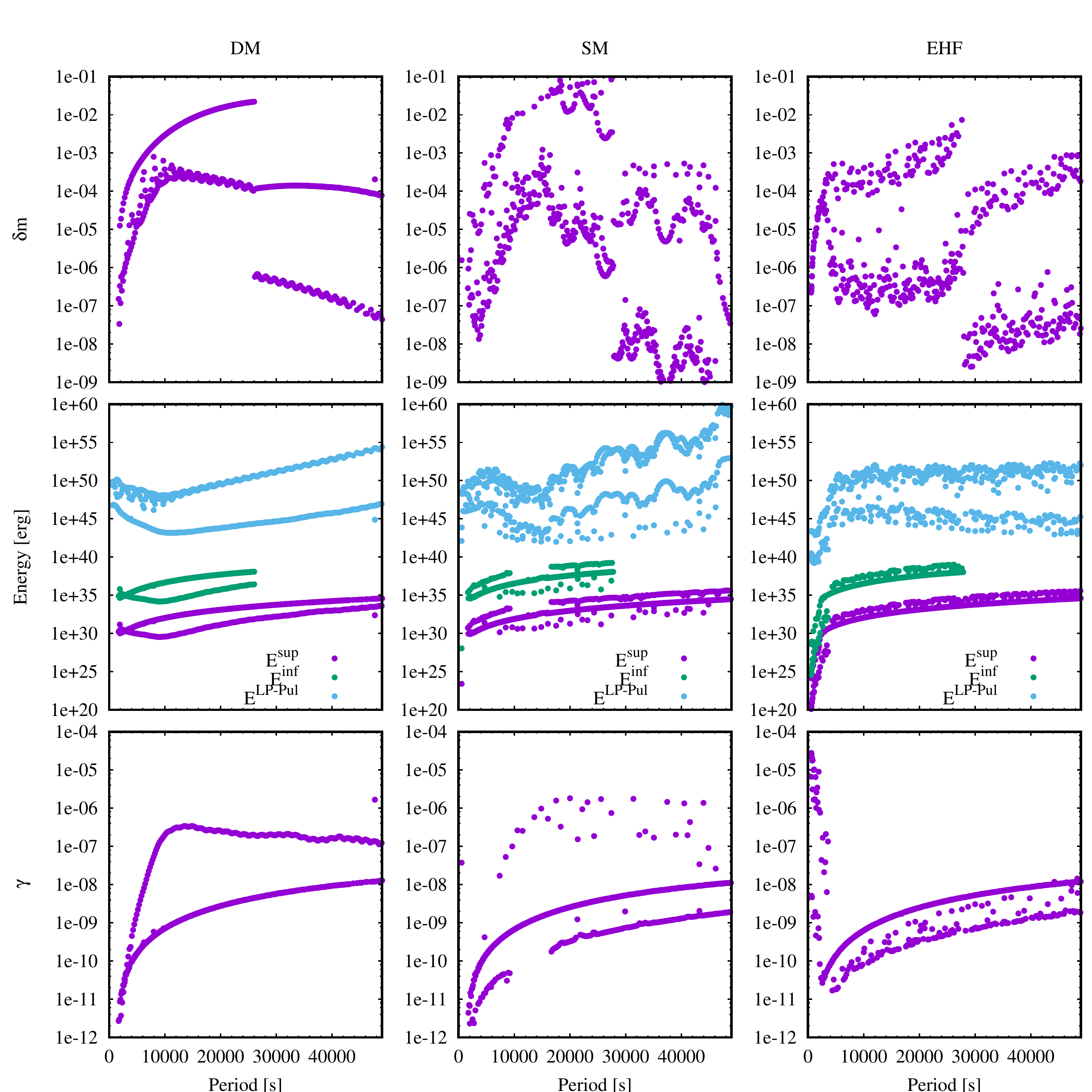}
\caption{{\bf Pulsations properties of DM, SM and EHF models with canonical initial abundances ($Y=0.285$)  during the first subflash and under the assumption of  $\eta=1$.} Rows and columns have the same meaning as in supplementary figure \ref{Supp_fig1}.}
 \label{Supp_fig11}
\end{figure}

\begin{figure}
\includegraphics[width=14cm]{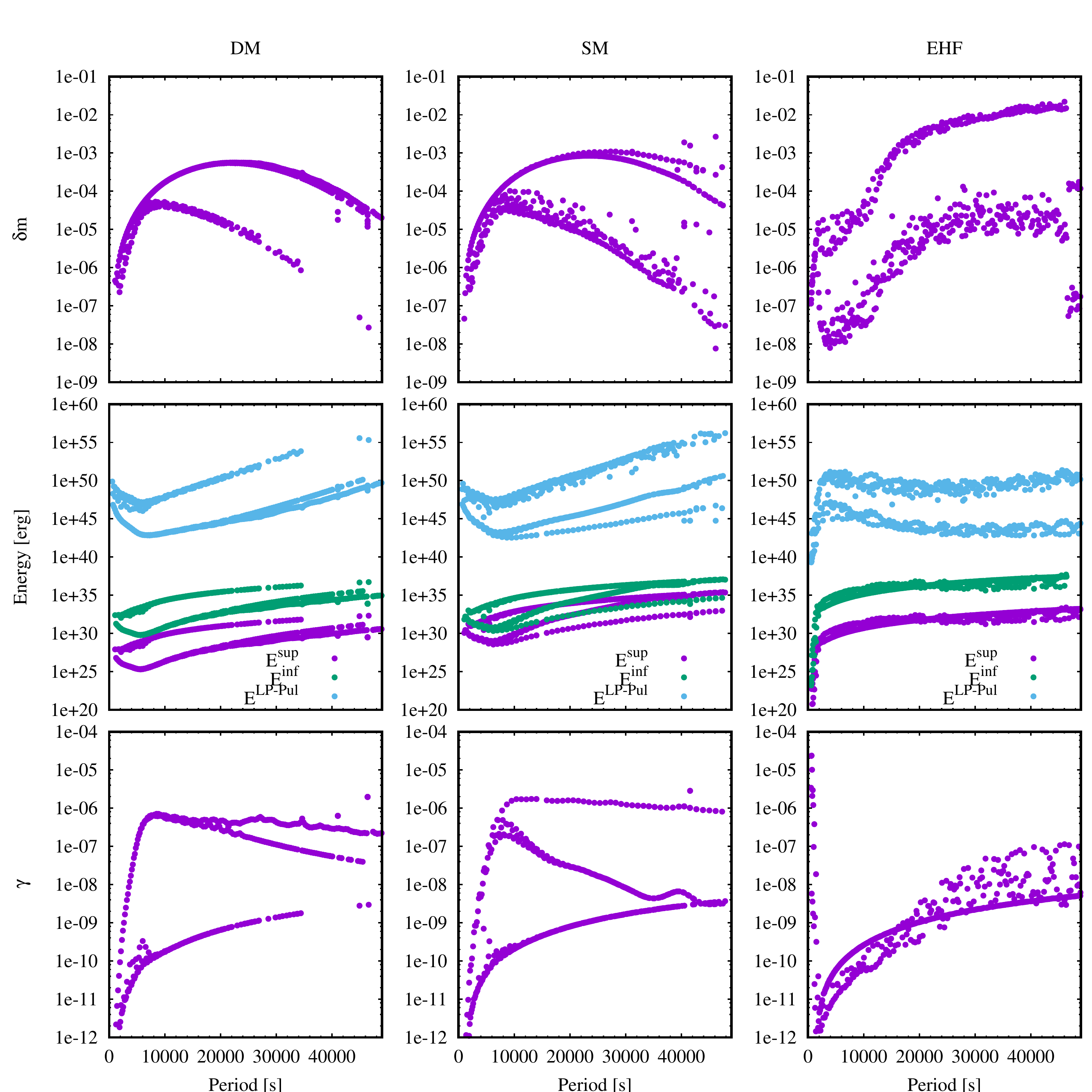}
\caption{{\bf Pulsations properties of DM, SM and EHF models with He-enriched  initial abundances ($Y=0.4$)  during the first subflash and under the assumption of $\eta=1$.} Rows and columns have the same meaning as in supplementary figure \ref{Supp_fig1}.}
 \label{Supp_fig22}
\end{figure}
\newpage
\subsection{Numerical Resolution.}
\begin{figure}
\centering
\includegraphics[width=9cm]{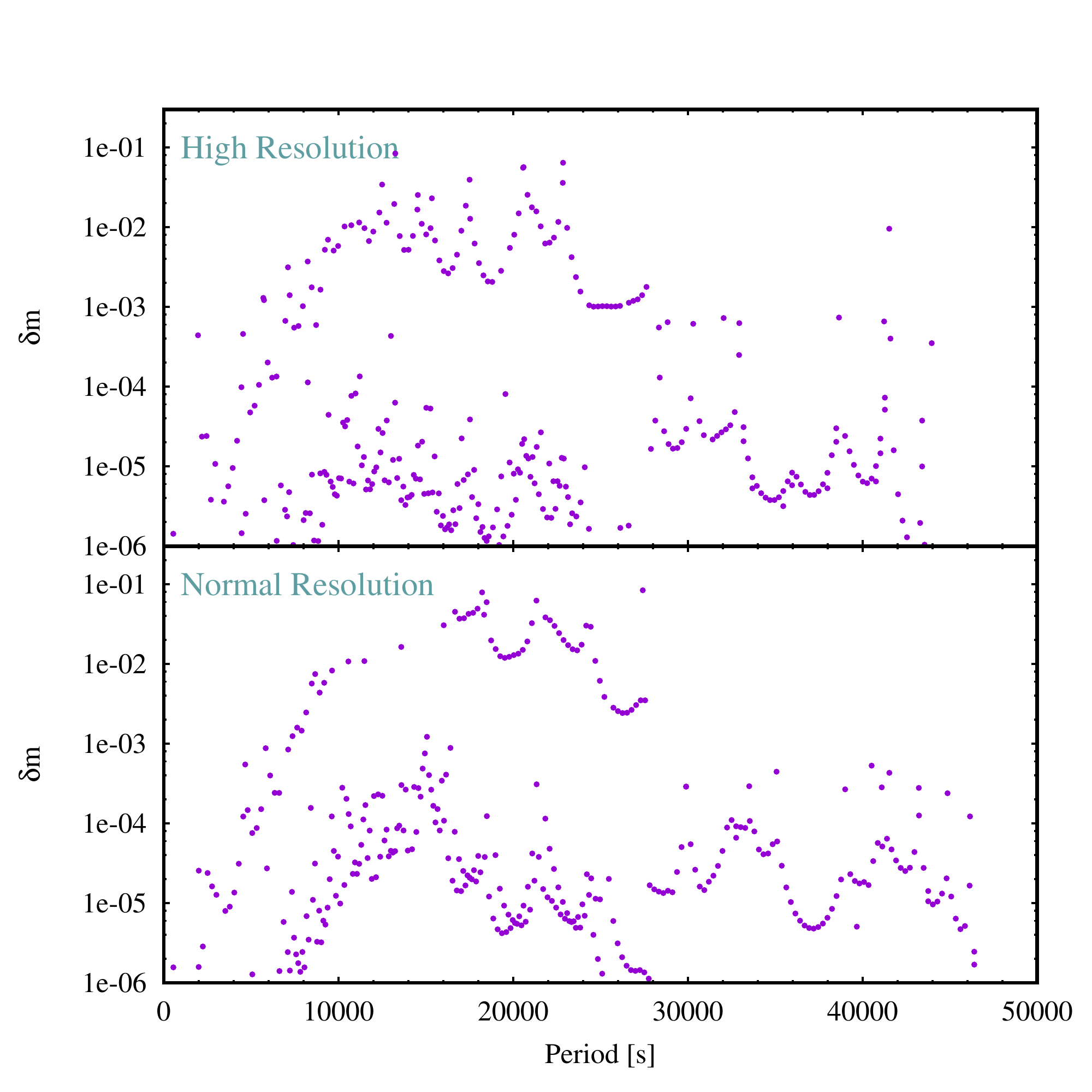}
\caption{{\bf Comparison of predicted amplitudes $\delta m$ for SM models computed with different mesh resolutions.} The high resolution model was computed with 64330 mesh points and compared with the 14569 mesh points in the normal resolution run.}
 \label{Supp_fig3}
\end{figure}

\begin{figure}
\centering
\includegraphics[width=9cm]{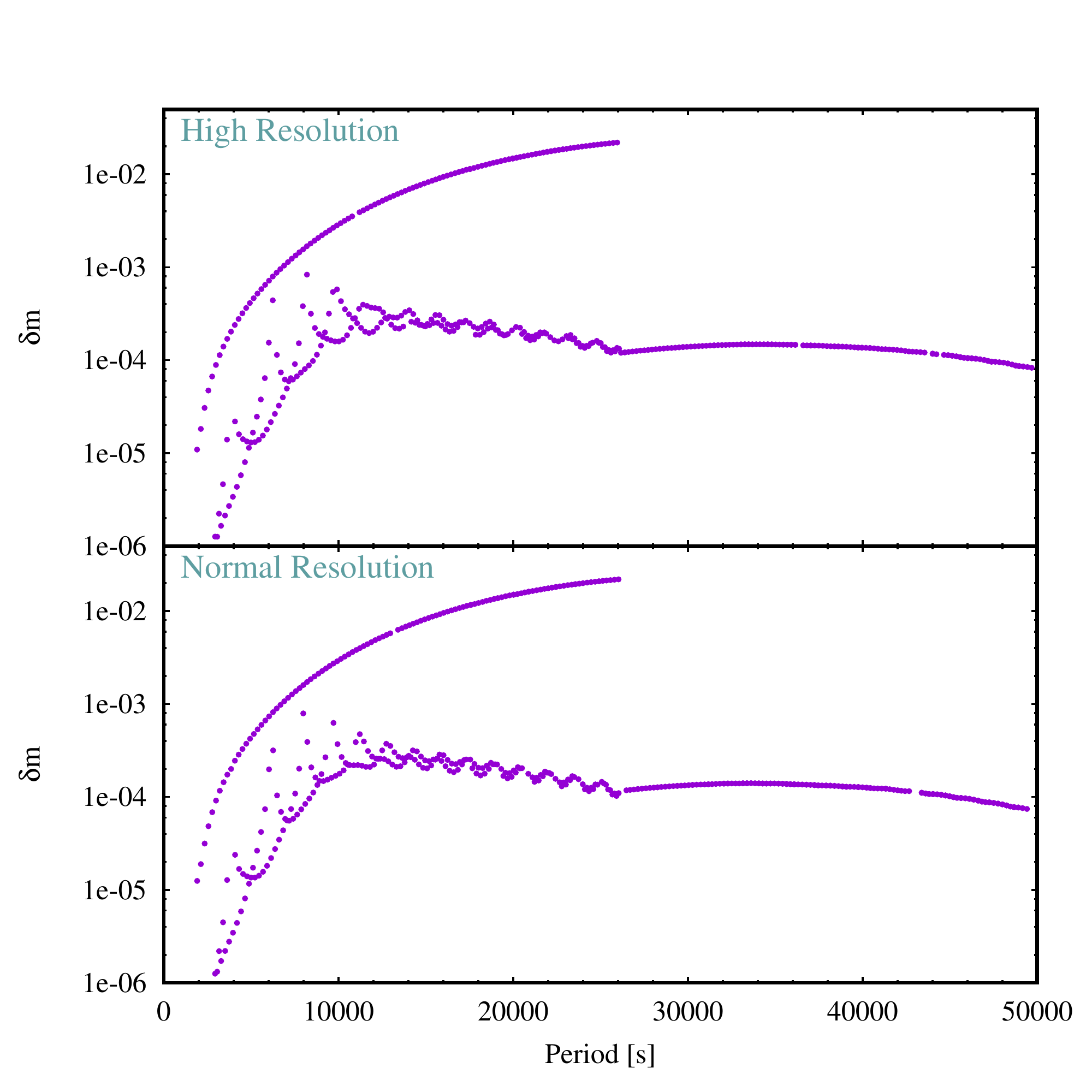}
\caption{{\bf Comparison of predicted amplitudes $\delta m$ for DM models computed with different mesh resolutions.} The high resolution model was computed with 44374 mesh points and compared with the 13629 mesh points in the normal resolution run.}
 \label{Supp_fig4}
\end{figure}

Spatial resolution is an important ingredient in this
computations. Due to the very large radial order of the low frequency
g modes a high spatial resolution is required. Modes with the longest
periods computed in this work have more than 700 nodes and very
complex eigenfunctions. For these reason all the models presented in
this work have high density spatial grids between 13000 and 15000 mesh
points. In addition we computed some sequences with extremely dense
grids to test the stability of our solutions with numerical
resolutions. Besides small differences due to the fact that the
background stellar evolution models change slightly due to the high
resolution grid, the qualitative results are very similar. High
resolution and normal resolution computations both show the same two
branches of low and high amplitude modes and a very similar behaviour
of amplitude with period (supplementary figures \ref{Supp_fig3} and
\ref{Supp_fig4}). This shows that eigenfuctions are well resolved
already at our normal resolution of about 13000 to 15000 mesh points.

\section{Acknowledgements}
This work was partially supported by ANPCyT through grant PICT
2016-0053, and by the MinCyT-DAAD bilateral cooperation program
through grant DA/16/07.  Funding for the Stellar Astrophysics Centre
is provided by The Danish National Research Foundation (Grant
DNRF106).  This research was supported in part by the National Science
Foundation under Grant No. NSF PHY-1748958. M.M.M.B. gratefully
acknowledges the finantial support by the Stellar Astrophysics Centre
(Denmark) that allowed him to participate in several Aarhus red giants
challenge workshops where the central ideas of this paper were
conceived.

\section{Author informations}

\subsection{Affiliations}

\noindent M. M. Miller Bertolami\\
Instituto de Astrof\'isica de La Plata, UNLP-CONICET, \\
Paseo del Bosque s/n, 1900 La Plata, Argentina; \\
e-mail: marcelo@mpa-garching.mpg.de\\ \\
T. Battich\\
Instituto de Astrof\'isica de La Plata, UNLP-CONICET, \\
Paseo del Bosque s/n, 1900 La Plata, Argentina; \\
e-mail: tbattich@fcaglp.fcaglp.unlp.edu.ar\\ \\
A. H. C\'orsico\\
Instituto de Astrof\'isica de La Plata, UNLP-CONICET, \\
Paseo del Bosque s/n, 1900 La Plata, Argentina; \\
e-mail: acorsico@fcaglp.fcaglp.unlp.edu.ar\\ \\
J. Christensen-Dalsgaard\\
Stellar Astrophysics Centre, Department of Physics and Astronomy, Aarhus University, \\
 Ny Munkegade 120, DK-8000 Aarhus C, Denmark; \\
Kavli Institute for Theoretical Physics, University of California Santa Barbara,\\
 CA 93106-4030, USA;\\
e-mail: jcd@phys.au.dk\\ \\
L. G. Althaus\\
Instituto de Astrof\'isica de La Plata, UNLP-CONICET, \\
Paseo del Bosque s/n, 1900 La Plata, Argentina; \\
e-mail: althaus@fcaglp.fcaglp.unlp.edu.ar\\ \\

\subsection{Contributions}

M.M.M.B. developed the idea, derived the theoretical expresions and
performed the pulsation computations. T.B. derived the theoretical
expresions and computed the stellar models with {\tt LPCODE}. A.H.G.
programed LP-PUL and discussed the modelling of stochastic
excitation.  J.C.D. provided insight in the nature of stochastic
oscillations and the modelling of stochastic excitation. All authors
participated in discussions of the results, their presentations in
figures and descriptions in the manuscript and in pinpointing the
conclusions.

\subsection{Competing interests}

The authors declare no competing financial interests.

\subsection{Corresponding author}

M. M. Miller Bertolami
\end{document}